\documentclass{article}

\usepackage{jheppub}

\usepackage{color}
\usepackage{amsmath}

\usepackage{ulem}

\newcommand{\roughly}[1]{\mathrel{\raise.3ex\hbox{$#1$\kern-0.85em
\lower1ex\hbox{$\sim$}}}}

\newcommand{\lsim}{\roughly<}

\def\exd{{\hbox{d}}}

\def\ba{\begin{eqnarray}}
\def\ea{\end{eqnarray}}
\def\be{\begin{equation}}
\def\ee{\end{equation}}

\def\bfk{{\bf k}}

\def\bfr{{\bf r}}

\def\ssA{{\scriptscriptstyle A}}
\def\ssB{{\scriptscriptstyle B}}
\def\ssD{{\scriptscriptstyle D}}
\def\ssF{{\scriptscriptstyle F}}
\def\ssE{{\scriptscriptstyle E}}

\def\ssV{{\scriptscriptstyle V}}

\def\EH{{EH}}
\def\DBI{{DBI}}

\def\cA{\mathcal{A}}
\def\cB{\mathcal{B}}

\def\cE{\mathcal{E}}
\def\cF{\mathcal{F}}
\def\cG{\mathcal{G}}

\def\cI{\mathcal{I}}
\def\cK{\mathcal{K}}

\def\cO{\mathcal{O}}
\def\cR{\mathcal{R}}

\def\cV{\mathcal{V}}

\def\mL{\mathfrak{L}}

\def\eff{{\rm eff}}

\def\nn{\nonumber}

\def\({\left(}
\def\){\right)}

\def\pref#1{(\ref{#1})}

\usepackage{graphicx}
\usepackage{dcolumn}
\usepackage{bm}
\usepackage{color}
\usepackage{ulem}

\title{Keeping an Eye on DBI: \\
Power-counting for small-$c_s$ Cosmology}

\author[a]{Ivana Babic,}
\author[a,b]{C.P.~Burgess}
\author[a,c,d]{and Ghazal Geshnizjani}
\affiliation[a]{Perimeter Institute for Theoretical Physics, Waterloo, Ontario N2L 2Y5, Canada }
\affiliation[b]{Physics \& Astronomy, McMaster University, Hamilton, Ontario L8S 4M1, Canada}
\affiliation[c]{Department of Applied Mathematics, University of Waterloo, Waterloo, Ontario N2L 3G1, Canada}
\affiliation[d]{Waterloo Centre for Astrophysics, University of Waterloo, Waterloo, Ontario N2L 3G1, Canada}

\date{}

\abstract{Inflationary mechanisms for generating primordial fluctuations ultimately compute them as the leading contributions in a derivative expansion, with corrections controlled by powers of derivatives like the Hubble scale over Planck mass: $H/M_p$. At face value this derivative expansion breaks down for models with a small sound speed, $c_s$, to the extent that $c_s \ll 1$ is obtained by having higher-derivative interactions like $\mL_\eff \sim (\partial \Phi)^4$ compete with lower-derivative propagation. This concern arises more generally for models whose lagrangian is given as a function $P(X)$ for $X = -\partial_\mu \Phi \partial^\mu \Phi$ --- including in particular DBI models for which $P(X) \propto \sqrt{1-kX}$ --- since these keep all orders in $\partial \Phi$ while dropping $\partial^n \Phi$ for $n > 1$. We here find a sensible power-counting scheme for DBI models that gives a controlled expansion in powers of three types of small parameters: $H/M_p$, slow-roll parameters (possibly) and $c_s \ll 1$. We do not find a similar expansion framework for generic small-$c_s$ or $P(X)$ models. Our power-counting result quantifies the theoretical error for any prediction (such as for inflationary correlation functions) by fixing the leading power of these small parameters that is dropped when not computing all graphs (such as by restricting to the classical approximation); a prerequisite for meaningful comparisons with observations. The new power-counting regime arises because small $c_s$ alters the kinematics of free fluctuations in a way that changes how interactions scale at low energies, in particular allowing $1-c_s$ to be larger than derivative-measuring quantities like $(H/M_p)^2$.  
}

\begin{document}

\maketitle
\section{Introduction}

Now that improved observations have made cosmology a precision science, theorists must raise their game when proposing predictions to be tested. This is true in particular for predictions about primordial fluctuations, whose measured near-scale-invariant pattern seems eerily consistent with those of quantum fluctuations \cite{quantumFluct} during a much-earlier accelerated epoch.

For want of alternatives, models for this earlier accelerated epoch are explored using semiclassical methods, which presuppose quantum fluctuations to be small correction to a leading classical result. In these times of precision cosmology the models of most interest are those for which such a semiclassical expansion is under control: {\it i.e.}~semiclassical corrections arise as a series in powers of known small dimensionless quantities. Such models are important because all practical predictions are necessarily approximate. Quantifying the small parameters that accompany whatever is {\it not} included in a calculation is what allows an assessment of the theoretical error associated with any given prediction; an assessment that is clearly a prerequisite for meaningful comparisons between theory and observations. 

Within the present state of the calculational art such control has proven possible for some inflationary models \cite{Inflation}, including the simple slow-roll models of most practical interest. For these models simple power-counting rules have been explicitly worked out\footnote{It would be very useful to have similar power-counting rules for non-inflationary models \cite{Alternatives} so that their theoretical errors could be equally well assessed.} \cite{InfPowerCount, ABSHclassic} using the framework of effective field theories (EFTs) \cite{EFTreview0} applied to gravity \cite{GREFT0} (see \cite{EFTreview, GREFT,EFTBook} for reviews using the same framework as used here). In these models semiclassical predictions arise as a controlled series in powers of\footnote{Fundamental units are used throughout, for which $\hbar = c = 1$.} $H/M_p$ and slow-roll parameters, where $H$ is the Hubble scale during inflation and $M_p$ is the reduced Planck mass.\footnote{That is: $M_p = (8 \pi G)^{-1/2}$ where $G$ denotes Newton's constant of universal gravitation.}  

The low-energy expansion in powers of $H/M_p$ is generic in cosmology. Low-energy expansions ultimately arise because any quantum treatment that includes gravity is non-renormalizable \cite{NRGR}, due to the fact that its coupling constant (Newton's constant $G$) has dimensions of inverse squared-mass. Consequently higher powers of $G$ necessarily combine with some other energy to provide a small dimensionless expansion parameter. Although the presence of ultraviolet divergences  mean this other energy scale could in principle be large, after renormalization it happens that the semiclassical (loop) expansion in cosmology is in powers of $GH^2 \propto (H/M_p)^2$ \cite{EFTreview0, GREFT0, EFTreview, GREFT, EFTBook}.

The purpose of this paper is to expand the class of models with a controlled semiclassical approximation to include those of the Dirac-Born-Infeld type \cite{DBI}, for which straightforward derivative expansions do not straightforwardly apply.

\subsection{The low-energy party line}

For the present purposes what is important about EFT arguments is that they imply that the lagrangian of the cosmological model of interest, say\footnote{Our metric is $(-+++)$ and we use Weinberg's curvature conventions \cite{GnC}, which differ from MTW conventions \cite{MTW} only in the overall sign of the definition of the Riemann tensor.}
\be \label{VanillaMod}
 \mL_{SR} = - \sqrt{-g} \, \left\{ \frac{M_p^2}{2} \, \Bigl[  \cR +  (\partial \hat \Phi)^2 \Bigr] + V(\hat \Phi) \right\} = - \sqrt{-g} \, \left[ \frac{M_p^2}{2} \, \cR + \frac12 \, (\partial \Phi)^2 + V(\Phi/M_p) \right] \,,
\ee
must always be regarded as just the leading part of a more general effective action, like 
\be \label{VanillaEFT}
   \mL_{EFT} = \mL_{SR} + c_1 \cR^2 + c_2 \cR^3 + c_3 (\partial \Phi)^4 + \cdots \,,
\ee
that involves the fewest (two or less) derivatives. In \pref{VanillaMod} $\cR$ denotes the spacetime metric's Ricci scalar, while in \pref{VanillaEFT} it more generically represents many terms that involve the Ricci and Riemann tensors and their derivatives, rather than just the Ricci scalar. The dimensionless and canonical fields $\hat \Phi$ and $\Phi$ are related by $\Phi = M_p \hat \Phi$. In principle $\mL_{EFT}$ contains an infinite series of terms, only a few of which are written explicitly in \pref{VanillaEFT}. On dimensional grounds a generic effective coupling --- like $c_2 \propto M^{-2}$ or $c_3 \propto M^{-4}$ above --- has dimensions of inverse powers of mass and the scale, $M$, appearing in this coupling is typically set by the lightest degrees of freedom that were integrated out in order to obtain the effective theory in the first place \cite{GREFT} (and so in particular is often much smaller than $M_p$). For instance $M$ might be the electron mass for applications to present-day cosmology while $M$ might be a UV scale like the string or Kaluza-Klein scale in an application to inflation that takes place at much higher energies.

When only a single scale (call it $H$) characterizes the low-energy physics of interest, a simple dimensional argument\footnote{Beyond counting dimensions, the power-counting formulae encountered below also keep track of powers of $4\pi$, which can sometimes be numerically important. See \cite{GREFT,EFTBook} for a more detailed description of how these factors are identified.} allows the determination of how any correlation function,\footnote{We follow the notation of \cite{ABSHclassic} and use $\cB$ to denote correlation functions like $\langle \phi \, \phi \rangle$, while reserving $\cA$ to denote amputated correlation functions, for which the Feynman rules for all external lines are removed.} $\cB$, depends on the scales $M$ and $M_p$ that appear within the effective couplings, $c_n$.  For instance, imagine computing a Feynman graph built using vertices determined by expanding about a classical background,
\be
  g_{\mu\nu} = \bar g_{\mu\nu} + \frac{2h_{\mu\nu}}{M_p} \quad \hbox{and} \quad \Phi = \varphi + \phi \,,
\ee
within the lagrangian $\mL_{EFT}$. Then a graph involving $L$ loops and $\cE$ external lines {\it etc.} leads to the a contribution to $\cB$ that is of order \cite{InfPowerCount, ABSHclassic}
\be \label{PCresult}
 \cB_\cE (H) \propto \frac{M_p^2}{H^2} \left( \frac{H^2}{M_p} \right)^\cE
 \left( \frac{H}{4 \pi \, M_p}
 \right)^{2L}
 \prod_{n \atop d_n = 0} \left[  \left( \frac{\mu^4}{H^2 M_p^2}
 \right) \right]^{V_n} 
 \prod_{n \atop d_n > 2} \left[ \left( \frac{H}{M_p}
 \right)^2 \left( \frac{H}{M}
 \right)^{d_n-4} \right]^{V_n}   \,. 
\ee
Here $n$ labels all possible types of interactions\footnote{`Interactions' here means cubic and higher monomials of the fluctuations $\phi$ and $h_{\mu\nu}$ that appear once $\mL_{EFT}$ is expanded about the classical background.} and $V_n$ denotes the number of times a vertex of a particular type appears in the graph of interest. $d_n$ denotes the number of derivatives in the vertex in question. This formula assumes the zero-derivative ($d_n = 0$) interaction are all taken from a potential of the form
\be \label{VvsPhi}
 V(\Phi) = \mu^4 \sum_{r=0}^\infty \lambda_r\left( \frac{\Phi}{M_p} \right)^r \,,
\ee
where $\lambda_r$ are dimensionless and at most of order unity (but for vanilla inflationary models are usually suppressed by slow-roll parameters). 

The dependence of \pref{PCresult} on $M$ and $M_p$ is fixed by how they appear in the lagrangian and the dependence on the low-energy scale $H$ is determined on dimensional grounds, assuming that ultraviolet divergences are regularized using dimensional regularization. It is crucial for this dimensional argument that only one low-energy scale exists, so in cosmological applications it is most useful for applications where all kinematic variables in $\cB$ are chosen to be of order the background Hubble scale $H$. 

Eq.~\pref{PCresult} shows that higher loops are small ({\it i.e.}~the semiclassical approximation is valid) if $H \ll 4\pi M_p$ and that interactions involving four or more derivatives are additionally suppressed provided $H \ll M$. Furthermore $d_n=0$ interactions do not give large contributions provided that $\mu^2 \lsim HM_p$ \cite{InfPowerCount} (which is in particular true when the scalar potential dominates the classical Friedmann equation, as usually assumed during inflation, since this implies $H^2 \sim V/M_p^2 \sim \mu^4/M_p^2$).   

The assumptions behind eq.~\pref{PCresult} apply in particular for correlation functions of fluctuations when evaluated at horizon exit $k/a = H$ during inflation, and reproduce standard inflationary results for quantities like $\cB_2 = \langle \phi \,\phi \rangle$ or $\langle h \, h \rangle$ and for $\cB_3 = \langle \phi \, \phi \, \phi \rangle$ or $\langle h \, h \, h \rangle$ and so on. For $\mu^2 \sim HM_p$ eq.~\pref{PCresult} predicts the dominant contributions arise when both of two conditions are satisfied: ($i$) $L = 0$ ({\it i.e.} only tree graphs), and $(ii)$ $V_n = 0$ for all interactions with more than two derivatives ({\it i.e.} $d_n > 2$). That is to say, the dominant contribution comes from purely classical physics involving the zero- and two-derivative terms (or classical physics using only the action \pref{VanillaMod}). Eq.~\pref{PCresult} states that this leading result is of order $\langle \phi\, \phi \rangle \sim H^2$. 

Even more usefully, eq.~\pref{PCresult} also identifies the origins of (and size) of subleading corrections. According to \pref{PCresult} higher loops ({\it i.e.} quantum effects) are suppressed by powers of $H/(4\pi M_p)$, while a higher derivative term like $(\partial \phi/M_p)^4$ first contributes at order\footnote{Notice that any graph involving a 4-point vertex and only two external lines must involve at least one loop.} $\langle \phi \,\phi \rangle \sim H^6/M_p^4$, and so on. Additional suppressions due to dimensionless slow-roll parameters, such as might be hidden within the dimensionless coefficients $\lambda_n$ of \pref{VvsPhi}, can be similarly followed \cite{ABSHclassic}.

\subsection{DBI models}

The purpose of this paper is to extend the above power-counting arguments to a broader class of inflationary models that do not admit as straightforward a derivative expansion. In particular the goal is to extend them to include the Dirac-Born-Infeld (DBI) form, for which a DBI scalar is described by a lagrangian density $\mL_\DBI = \sqrt{-g} \, L(X,\Phi)$ where \cite{DBI} 
\be \label{pravi DBI}
    L(X, \Phi) = -\frac{1}{f(\Phi)}   \sqrt{1-f(\Phi)X}-W(\Phi)  \,,
\ee
and (again) $\Phi$ is a scalar field while\footnote{The signs are chosen so that $X = \dot\Phi^2 \ge 0$ for homogeneous functions of time.} 
\be
   X = -(\partial \Phi)^{2} = - g^{\mu\nu} \partial_\mu \Phi \, \partial_\nu \Phi \,. 
\ee
The model is specified once the functions $W(\Phi)$ and $f(\Phi)$ are given, with the function $f(\Phi)$ not passing through zero (and in practice chosen to be positive).

What makes the DBI model --- and more general models based on other forms\footnote{See \cite{JJKT} for a review with references of these and other cosmological models.} for $L(X,\Phi)$ \cite{Kflation} unusual from the above power-counting point of view is the fact that its action keeps all orders in single derivatives like $X$ while ignoring all higher derivatives like $\partial_\mu \partial_\nu \Phi$. This clearly requires going beyond the usual derivative expansion that underpins control of approximations within the simplest slow-roll models (see \cite{Burgess:2014lwa, Trodden} for a discussion of higher-derivative effective couplings and their implications). 

\subsubsection*{Properties of the inflationary scenario}

It is useful to summarize briefly some properties of inflationary models built using the DBI action of \pref{pravi DBI}, some of which are summarized here. For inflationary applications one writes
\be \label{WvsV}
  W(\Phi) = V(\Phi) - \frac{1}{f(\Phi)} \,,
\ee
so that when the product $f(\Phi) X$ is small the DBI lagrangian density is well-approximated by
\be
  \mL_{DBI}(X, \Phi) \simeq \sqrt{-g} \, \left[\frac12\, X - V(\Phi) \right] \,.
\ee
This reveals $V(\Phi)$ to be the potential for a canonically normalized scalar in this regime.   

The square-root action resembles the action of a relativistic particle, for which $f(\Phi) X \to V^2$ would be the squared particle speed. This analogy motivates defining
\be \label{gamma}
    \gamma := \frac{1}{\sqrt{1-f(\Phi)X}} \,,
\ee
which is analogous to the usual relativistic time-dilation factor $\gamma = 1/\sqrt{1 - V^2}$ that satisfies $\gamma \to 1$ in the nonrelativistic regime ($V^2 \to 0$) and $\gamma \to \infty$ in the ultrarelativistic regime ($V^2 \to 1$).  It is this relativistic particle analogy that suggests that the limit $V \to 1$ with $\dot V$ small should make sense.

For DBI models the relativistic regime $\gamma \gg 1$ is also novel because in it, the wave speed for $\Phi$ fluctuations --- {\it i.e.} the cosmic fluid's speed of sound --- turns out to be given by
\be \label{csvsgamma}
    c_{s}^{2} = \frac{\partial p/\partial X}{\partial \rho/\partial X} = \frac{\partial L/\partial X}{\partial L/\partial X + 2X \partial^2 L/\partial X^2} = \frac{1}{\gamma^{2}} \,,
\ee
where the last equality specializes to the DBI lagrangian \pref{pravi DBI}. For DBI inflation small $c_s$ (compared to the speed of light) necessarily goes hand-in-hand with the regime $\gamma \gg 1$.

The gravitational response relevant to inflation is found by minimally coupling the DBI scalar to the Einstein-Hilbert action for gravity, $\mL = \mL_\EH + \mL_{DBI}$, where (as usual)
\be
  \mL_\EH = - \sqrt{-g} \; \left(  \frac{M_p^2}{2} \, \cR \right) \,.
\ee
The stress-energy tensor appearing in the Einstein equations is then given by 
\be
    T^{\mu \nu} := \frac{2}{\sqrt{-g}} \, \frac{\delta S_\DBI}{\delta g_{\mu\nu}} = (p+\rho)u^{\mu}u^{\nu} + p \,g^{\mu \nu} \,,
\ee
where  
\be 
    \rho(X,\Phi)  := 2X\frac{\partial L}{\partial X} - L(X, \Phi) 
     \,, \qquad    
     p(X, \Phi) := L(X,\Phi) \quad \hbox{and} \quad 
    u_{\mu} := \frac{\partial_\mu \Phi}{\sqrt{X}}  \,.
\ee
This stress-energy takes the form of a perfect fluid when specialized to spatially homogeneous configurations, for which $\Phi = \varphi(t)$ since in this limit $u^\mu$ becomes a time-like 4-velocity (see \cite{BPS} for a discussion of constraints on these types of scalar-based preferred-frame models).  In this case $\rho$ and $p$ are the fluid's energy density and pressure, which in the particular case of the DBI model become
\be 
    \rho(X,\Phi)  = \frac{\gamma}{f(\Phi )} + W(\Phi) \,,
\ee
with $\gamma$ as defined above.

The conditions for a homogeneous configuration, $\Phi = \varphi(t)$, to inflate are then found using the Friedmann equation
\be
    H^{2} = \frac{1}{3 M_{p}^{2}} \left[\frac{\gamma}{f(\varphi )} + W(\varphi)\right] =  \frac{1}{3 M_{p}^{2}} \left[\frac{\gamma-1}{f(\varphi )} + V(\varphi)\right] \,,
\end{equation}
where the second equality uses \pref{WvsV}. Differentiating, using the Einstein equation $\dot \rho + 3H(\rho + p) = 0$, the first slow-roll parameter becomes
\be 
    - \frac{\dot H}{H^2} = \frac{3(\rho + p)}{2\rho}
    = \frac{3}{2}\left[ \frac{\gamma- \gamma^{-1}}{\gamma  + f(\Phi)W(\varphi)}\right] 
    = \frac{3}{2}\left[ \frac{\gamma- \gamma^{-1}}{\gamma -1  + f(\varphi)V(\varphi)}\right] \,.
\end{equation}
For slowly moving fields, $f(\varphi) X \ll 1$, this is small whenever the usual slow-roll condition $X \ll V$ is satisfied. The main interest in these models comes from the limit $\gamma \to \infty$, however, for which $- \dot H/H^2 \ll 1$ is ensured so long as
\be
    1 \ll \gamma \ll f(\varphi) V(\varphi) \,,
\ee
which is novel because it does not require $V$ to have a small derivative. In this regime the Friedmann equation simplifies to 
\be \label{HiV}
    H^{2} \simeq \frac{ V(\varphi)}{3 M_{p}^{2}} \,.
\ee
For later purposes, what is important is that no powers of $c_{s}$ hide inside the Hubble parameter.

There is one final noteworthy feature of the $\gamma \gg 1$ regime. Homogeneity implies $\Phi = \varphi(t)$, while the field equations for the lagrangian \pref{pravi DBI} are
\be
  \frac{1}{[1 - f(\varphi) \dot\varphi^2]^{3/2}} \left[ \ddot \varphi - \frac{f'(\varphi)}{f^2(\varphi) } \left(1 - \frac32 \, f(\varphi) \dot \varphi^2 \right) \right]  + W'(\varphi) = 0   \,,
\ee
which implies $\ddot \varphi$ is given by 
\be
 \ddot \varphi = \frac{f'(\varphi)}{f^2(\varphi) } \left(1 - \frac32 \, f(\varphi) \dot \varphi^2 \right)  - W'(\varphi) \Bigl[ 1- f(\varphi) \dot \varphi^2 \Bigr]^{3/2} = -  \frac{f'(\varphi)}{2f^2(\varphi) } \left(1 - \frac3{\gamma^2}  + \frac{2}{\gamma^3} \right)  - \frac{V'(\varphi)}{\gamma^3} \,,
\ee
and the second equality trades $\dot\varphi$ for $\gamma$ using $f \dot \varphi^2 = 1 - (1/\gamma)^2$. The noteworthy feature of this equation is that $\ddot \varphi \to 0$ for $\gamma \to \infty$, provided only that $f'/f^2$ is chosen to be sufficiently small (a slow-roll condition) largely independent of the steepness of $V$. Consequently second and higher derivatives of $\varphi(t)$ naturally tend to zero as $\gamma \to \infty$ even though $f(\varphi) \dot\varphi^2$ approaches unity.

\subsubsection*{DBI EFT}

The logic of the rest of this paper is to adapt the power-counting arguments of \cite{InfPowerCount, ABSHclassic} to the DBI model defined by \pref{pravi DBI}. The first step when doing so is to extend \pref{pravi DBI} into an EFT framework, which means extending it to include a broader class of effective operators including those with higher dimensions than \pref{pravi DBI}. We call this extension DBI EFT to distinguish it both from the DBI model \pref{pravi DBI} and from a generic EFT that includes all possible operators. 

For DBI inflation, the simplest case is to assume only a single scalar field, $\Phi$, is present at low energy, coupled both to itself and to the spacetime metric. In what follows, we define a DBI EFT as one where the field's Wilsonian effective lagrangian contains a piece of the form $\mL_{\rm eff} = \sqrt{-g} \;  L_{\rm eff}$, with
\be \label{L0def}
 L_{\rm eff} = - \sqrt{1 -  \cG(\Phi) X} \;\hat{\cF}(\Phi, \partial \Phi, \partial^2 \Phi, \dots)   \,,
\ee
where  
%
the positive function $\cG(\Phi)$ depends on $\Phi$ but not its derivatives.\footnote{Strictly speaking, if there is only the one field $\Phi$ then the function $\cG$ can be set to unity by performing a field redefinition to the new field $\widetilde\Phi$ satisfying $\exd\widetilde\Phi = \sqrt{\cG(\Phi)} \; \exd \Phi$. This redefinition is not performed here, both to connect better to the DBI model, and because a similar redefinition is not possible when there are multiple fields.} The square-root term of the vanilla DBI model given in \pref{pravi DBI} corresponds to the special cases $\cG(\Phi) = f(\Phi)$ and $\hat{\cF} = 1/f(\Phi)$.

The new feature relative to \pref{pravi DBI} is the function $\cF$, which is now allowed to depend on $\Phi$ and all of its derivatives in a local, but otherwise arbitrary, way. We can rewrite this function in terms of dimensionless variables such that,  
%
%
\be \label{L0def2}
   \hat{\cF}(\Phi, \partial \Phi, \partial^2 \Phi, \dots) = \cF \left(\frac{\Phi}{v}, \frac{\partial \Phi}{Mv}, \frac{\partial^2 \Phi}{M^2 v}\,, \cdots \right) \,,
\ee
where $v$ is the large scale in the EFT, against which $\Phi$ is compared --- {\it e.g.}~in the lagrangian \pref{VanillaMod}), $v \sim M_p$ and $M$ is assumed  to be the corresponding quantity for derivatives.

The defining feature of the DBI EFT is the appearance of the square-root factor in front of the entire lagrangian in $\mathfrak{L}_{\rm eff}$, a feature that can sometimes be justified as a natural consequence of symmetries in higher dimensions \cite{DBIsym, DBIsym2}. Indeed, in the original DBI models the fields $\Phi^a$ are regarded as being the low-energy Goldstone modes for the breaking of spacetime symmetries by D-brane position fields within a string-theory framework. These often require the lagrangian to transform like a scalar density with respect to spacetime and target space coordinate redefinitions, leading to lagrangians of the form $\mathfrak{L} = \sqrt{-\gamma} \;L$. Here, $\gamma := \hbox{\rm det} \, \gamma_{\mu\nu}$ with
$\gamma_{\mu\nu}$ an appropriate field-dependent metric, such as
\be \label{inducedGab}
  \gamma_{\mu\nu} := g_{\mu\nu} + \partial_\mu \Phi^a \partial_\nu \Phi^b\, \cG_{ab}(\Phi) \,.
\ee
In this relation,  $\cG_{ab}$ denotes the target-space metric. As can be shown by explicit evaluation, for a single scalar field\footnote{A similar simplification occurs for multiple fields $\Phi^a$ depending only on a single independent coordinate.} $\Phi$, the determinant of the metric \pref{inducedGab} satisfies
\be
  \sqrt{-\gamma} = \sqrt{1 - \cG(\Phi) X} \,,
\ee
thereby providing the overall square-root factor of \pref{L0def}. 

In following sections, we develop power-counting rules under the assumption that the action has this square root factor, and do not worry too much about origin of the symmetries (although we do expect that our success in setting up a sensible EFT framework, likely has its roots in such symmetry arguments). We return to the inclusion of terms not involving the square root --- such as the scalar-potential term in \pref{pravi DBI} or the Einstein-Hilbert lagrangian itself --- when power-counting with gravity in \S\ref{sec:PCDBIgrav}.

\bigskip
\subsection{The main point}

The additional step in order to apply power-counting to this theory, relative to \cite{InfPowerCount, ABSHclassic} is to keep track of how amplitudes depend on the new small parameter $c_s \ll 1$. This existence of the small parameter $c_s$ is important because it turns out to suppress all interactions in the DBI EFT {\it except} those that come from the expansion of the $\sqrt{1-\cG X}$ factor of the action. This is why it can be self-consistent to keep the interactions from $\mL_\DBI$ within the classical approximation, while neglecting terms with similar numbers of derivatives appearing in $\cF$.

One must still check that non-classical contributions (graphs involving loops) are also consistently organized in powers of $c_s$, and this occurs because for small $c_s$ the kinematics of fluctuations is significantly different from those of standard relativistic modes (because they are no longer relativistic), and this in turn changes how effective interactions scale at low energies. This investigation is the topic of the following sections, which show explicitly how successive Feynman graphs depend on $c_s$, once expanded about a $c_s$-dependent background configuration.

The detailed discussion starts in \S\ref{sec:PCDBI} with just the DBI scalar on a flat space-time background, temporarily ignoring Hubble expansion and metric fluctuations. This is then followed up in \S\ref{sec:PCDBIgrav} with a discussion that includes background evolution and metric fluctuations and how these change the power-counting arguments of \S\ref{sec:PCDBI}. A final summary and discussion is given in \S\ref{sec:Conclusions}. 

\section{Power-counting (scalars only)}
\label{sec:PCDBI}

To warm up, this section describes power-counting just for the square-root action of DBI scalar, putting off until later any discussion of other interactions (like the coupling to gravity and the potential term $W(\Phi)$). This allows the main issues to be exposed in a slightly simpler setting before passing on to the general case. To this end, for now we only consider the effective action given in \pref{L0def} and \pref{L0def2}, repeated here for convenience:
\be \label{L0defrpt}
 \mathfrak{L}_{\rm eff} = - \sqrt{1 - \cG(\Phi) X} \;\; \cF  \left(\frac{\Phi}{v}, \frac{\partial \Phi}{Mv}, \frac{\partial^2 \Phi}{M^2 v} \right)  
 \,,
\ee
where restricting to flat spacetime implies
\be
  X = -\eta^{\mu\nu} \partial_\mu \Phi \, \partial_\nu \Phi  = \dot\Phi^2 - (\nabla \Phi)^2 \,.
\ee

\subsection{Vanilla DBI warm-up}

In order to appreciate the the special role played by the square-root term in \pref{L0defrpt}, it is instructive first to study the simplest case, where $\cG = 1/\Lambda^4$ and $\cF = \Lambda^4$ are both $\Phi$-independent: 
\begin{equation}\label{eqn::DBI_action}
	S_\DBI = - \Lambda^4  \int \exd t\, \exd^3x\;\sqrt{1-\frac{X}{\Lambda^4}}  =  \int \exd t\, \exd^3x\; \left[- \Lambda^4 + \frac{X}2 + \frac{X^2}{8 \Lambda^4} + \cdots \right] \,,
\end{equation}
with $\Lambda^2 = Mv$ defined for notational convenience. Here the second equality gives the action's expansion in the small-derivative limit ($|X| \ll \Lambda^4$, in which case $c_s \simeq 1$), however the main point in the discussions below is to systematize expansions about homogeneous background solutions for which $X/\Lambda^4 = \dot \varphi^2/\Lambda^4$ is {\it not} small, since this is the $c_s\ll 1$ regime. 

 The field equation obtained by varying $\Phi$ is 
\be \label{homogFE}
   \partial_\mu \left( \frac{\partial^\mu\Phi  }{\sqrt{1 - X/\Lambda^4}} \right)  = 0 \,,
\ee
which for homogeneous configurations becomes
\be \label{homogFE2}
  \frac{\exd}{\exd t} \left( \frac{ \dot{\varphi}  }{\sqrt{1 - \dot{\varphi}^2/\Lambda^4}} \right)  = 0 \quad \implies \quad \frac{ \dot{\varphi}  }{\sqrt{1 - \dot{\varphi}^2/\Lambda^4}}  = \hbox{const.} \,,
\ee
with solutions
\be \label{varphidotisconstant}
  \varphi = \varphi_0 + \dot\varphi_0 \, t \,,
\ee
for constants $\varphi_0$ and $\dot\varphi_0$. 

The goal is to characterize the dynamics of quantum and classical fluctuations of $\Phi$ about a general spatially homogeneous background-field configuration, $\varphi(t)$, that depends only on time:
\begin{equation}\label{eqn::background_perturbation_split}
	\Phi(t,\bfr) = \varphi(t)+\phi(t,\bfr) \,.
\end{equation}
Although $\phi$ and its derivatives are assumed small compared to background values $\varphi$ and $\dot{\varphi}$,\footnote{See \cite{SmallcsPC} for a classical discussion of the domain of validity of perturbative methods in small-$c_s$ models.} it is {\it not} assumed that $\dot \varphi^2/\Lambda^4$ is small. Inserting \pref{eqn::background_perturbation_split} into the action \eqref{eqn::DBI_action} gives
\ba\label{eqn::perturbation_split_action}
	S_\DBI &=&- \Lambda^4  \int \exd t \,\exd^3x\; \sqrt{ \left( 1 -\frac{\dot{\varphi}^2}{\Lambda^4} \right) - \left(\frac{2\dot{\varphi}}{\Lambda^4} \right) \dot \phi -\frac{\dot{\phi}^2}{\Lambda^4}+\frac{(\nabla \phi)^2}{\Lambda^4}} \nn\\
	&=& - \int \exd t \,\exd^3x \; \sqrt{1-\frac{\dot{\varphi}^2}{\Lambda^4}} \left[ \Lambda^4 + \frac{(\nabla \phi)^2}{2(1-\dot\varphi^2/\Lambda^4)} -  \frac{\dot \phi^2}{2(1-\dot\varphi^2/\Lambda^4)^2} + \cdots \right] \,,
\ea
where the ellipses represent terms involving at least three powers of $\phi$. Linear terms in $\phi$ are absent in the second line of \pref{eqn::perturbation_split_action} by virtue of the background field equation, \pref{homogFE}, satisfied by $\varphi$. 

Defining
\begin{equation}\label{eqn::DBI_sound_speed}
 	c_{s} := \sqrt{ 1-\frac{\dot{\varphi}^2}{\Lambda^4}} \,,
 \end{equation}
the second line of eq.~\pref{eqn::perturbation_split_action} is conveniently rewritten
\be\label{eqn::perturbation_split_action2}
	S_\DBI = \int \exd t \,\exd^3x \;  \left\{ - c_s \Lambda^4 + \frac{1}{2c_s^3} \Bigl[\dot\phi^2 -c_s^2 (\nabla \phi)^2 \Bigr] + \cdots \right\} \,,
\ee
revealing $c_s$ to be the propagation speed of linearized fluctuations. Notice that \pref{eqn::DBI_sound_speed} agrees with the result for this speed given in \pref{csvsgamma}, and that $c_s$ is time-independent for the solutions \pref{varphidotisconstant}.  The regime $c_s \ll 1$ clearly corresponds to $\dot \varphi^2 \lsim \Lambda^4$.

For later power-counting arguments it is convenient to canonically normalize the fluctuation field, $\phi$, and to rescale the time coordinate according to 
\be
  t \rightarrow t^{\prime} := c_s t
\ee
to keep space and time derivatives on the same footing. Using this temporal rescaling and defining the canonical field by 
\be
  \phi \rightarrow \psi := \frac{\phi}{c_s} \,,
\ee
the action \pref{eqn::perturbation_split_action2} then becomes $c_s$-independent,
\begin{equation}\label{eqn::time_rescaled_quadratic_action}
	S^{(2)} = \int \exd t^{\prime} \exd^3x \left\{ - \Lambda^4 + \frac12 \biggl[ (\psi')^{2} - (\nabla {\psi})^2 \bigg] \right\} \,,
\end{equation}
where primes denote differentiation with respect to $t'$. The point of using these variables is that they are $c_s$-independent at the saddle point that dominates the path integral in a Gaussian perturbative expansion. 

In terms of these variables the first line of eq.~\pref{eqn::perturbation_split_action} becomes
\ba \label{DBIrescaled}
	S_\DBI 
	&=& -\int \exd t' \,\exd^3x\; \Lambda^4  \sqrt{1 - 2\sqrt{1-c^2_s} \; \frac{{\psi}^{\prime}}{\Lambda^2} - c^2_s\bigg(\frac{{\psi}^{\prime}}{\Lambda^2}\bigg)^2+ \frac{(\nabla{{\psi}})^2}{\Lambda^4}} \\
	&=& -\int \exd t' \,\exd^3x\; \Lambda^4  \sqrt{1 -  \frac{2{\psi}^{\prime}}{\Lambda^2} + \frac{(\nabla{{\psi}})^2}{\Lambda^4}} \Bigl[ 1 + \cO(c_s^2) \Bigr] \,,\nn
\ea
which inverts \pref{eqn::DBI_sound_speed} to write $\dot\varphi = \Lambda^2 \sqrt{1-c_s^2}$. As the last equality emphasizes, these new variables show that the action has a nonsingular limit as $c_s \to 0$ with $\psi$ and $t'$ held fixed, allowing the DBI action to be written
\be \label{eqn::sqrt_expansion3}
- \frac{\Lambda^4}{c_s}  \int \exd t^\prime \, \exd^3x\;  \sqrt{1-\frac{X}{\Lambda^4}}  =
	- \Lambda^4  \int \exd t^\prime \, \exd^3x\;  \sum_{n} \mathfrak{a}_{n}(c_s) \mathcal{O}_n\bigg(\frac{{\psi}^{\prime}}{\Lambda^2} \,, \frac{\nabla{\psi}}{\Lambda^{2}}\bigg),
\ee
where the sum is over rotationally invariant monomials in $\psi'$ and $\nabla \psi$. The dimensionless coefficients 
\be
   \mathfrak{a}_{n}(c_s) = \sum_{k=0}^\infty \mathfrak{a}_{nk} c_s^{2k}
\ee
are functions of $c_s$ that remain nonsingular (and generically nonzero) as $c_s \to 0$. As is clear from the first line of \pref{DBIrescaled} all terms involving nontrivial powers of $c_s$ also have at least one factor of the time-differentiated fluctuation field, $\psi'$. 

For later use the leading few coefficients $\mathfrak{a}_n$ can be read off from the explicit expansion of \pref{DBIrescaled} in powers of the fields. Keeping in mind \pref{eqn::DBI_sound_speed}, expanding \pref{DBIrescaled} out to four powers of $\psi$ leads to
\ba \label{DBIpsiexpn}
 - \frac{\Lambda^4}{c_s} \sqrt{1-\frac{X}{\Lambda^2}}  &=& -\Lambda^4 + \Lambda^2 \sqrt{1-c_s^2} \;\psi' + \frac12 \, (\psi')^2 +  \sqrt{1-c_s^2} \,\frac{(\psi')^3}{2\Lambda^2} + \left( \frac54 - c_s^2 \right) \frac{(\psi')^4}{2\Lambda^4} \\
  && \quad -  \left[1 + \sqrt{1-c_s^2} \; \frac{ \psi'}{\Lambda^2}  - (3- 2 c_s^2)\frac{ (\psi')^2}{2\Lambda^4} \right] \frac12(\nabla \psi)^2  + \frac{ (\nabla \psi \cdot \nabla \psi)^2}{8\Lambda^4} + \cdots \,,\nn
\ea
where ellipses now include at least five factors of $\psi$ (all of which are differentiated). 
\subsubsection*{Contributions from $\cF$}

Let's now return to the general lagrangian of \pref{L0defrpt}, with $\cF$ not constant. In particular, writing $\cF$ as a sum over effective interactions allows it to be written in terms of the rescaled variables $t'$ and $\psi$,
\be \label{cFexpn}
	\cF = v^2 M^2 \sum_n \mathfrak{b}_n \cO_n \left( \frac{\phi}{v} \,, \frac{\dot\phi}{Mv} \,, \frac{\nabla \phi}{Mv} \,, \cdots \right) 
	= v^2 M^2 \sum_n \mathfrak{b}_n \cO_n \left( \frac{c_s\psi}{v} \,, \frac{c_s^2 \psi'}{Mv} \,, \frac{c_s \nabla \psi}{Mv} \,, \cdots \right) \,,
\ee
showing that each factor of $\psi$ and $\partial_{t'}$ appearing in $\cF$ comes with a factor of $c_s$. Interactions coming from $\cF$ within $\mathfrak{L}_{\rm eff}$ obtained by expanding about a saddle point for the DBI action are therefore always suppressed by factors of $c_s$ in the small-$c_s$ limit.  The first terms in the derivative expansion for $\cF$ are
\ba  \label{cFpsiexpn}
  \cF &=& \cF_{0} + \cF_{1} \, \frac{\dot \phi}{Mv} + \cF_{2} \, \left(\frac{\dot \phi}{Mv}\right)^2 + \cF_{3} \,  \frac{(\nabla \phi)^2}{2M^2v^2} + \cF_{4} \, \frac{\ddot \phi}{M^2v} +  \cdots \nn\\
  &=& \cF_{0} + \cF_{1} \, \frac{c_s(c_s \psi)'}{Mv} + \cF_{2}\, \left[ \frac{c_s(c_s \psi)'}{Mv} \right]^2 + \cF_{3}  \, \frac{(c_s \nabla \psi)^2}{2M^2v^2} +  \cF_{4}  \, \frac{c_s [c_s(c_s \psi)']'}{M^2v}  +  \cdots \,,
\ea
where each of the coefficient functions, $\cF_k$, potentially are functions of $\phi/v = c_s \psi/v$, without derivatives. Factors of $c_s$ are kept within the time derivatives because the presence of $\cF$ can cause $\dot\varphi$ (and so also $c_s$) to evolve with time. For inflationary applications this evolution is assumed to be subject to slow-roll conditions. If a field redefinition is not performed to ensure $\cG$ is a constant, its contributions to the lagrangian's interaction terms are found in a similar way.

Combining the expansions of $\cF$ and of the factor $\sqrt{1- X/\Lambda^4}$ given above --- for $\Lambda^2 = Mv$ --- the complete scalar effective Lagrangian for the canonical perturbation, $\psi$, becomes
\be\label{eqn::effective_Lagrangian}
	 \mathfrak{L}_{\rm eff} = M^2 v^2 \sum_n \mathfrak{a}_n(c_s) \,\cO_n \left( \frac{\psi}{v} \,, \frac{\partial_{t'}}{M} \,, \frac{\nabla}{M}  \right) \sum_m \mathfrak{b}_m \cO_m \left( \frac{c_s\psi}{v} \,, \frac{c_s\partial_{t'}}{M} \,, \frac{\nabla}{M} \right) \,,
\ee
where all possible local combinations of fields and their derivatives are to be included. The products of operators $\cO_n$ and $\cO_m$ can be re-expanded in terms of the basis of interaction terms and the new coefficients have the $c_s$-dependence obtained by summing over the relevant coefficients $\mathfrak{a}_n$ and $\mathfrak{b}_m$. In what follows it often proves useful to label interactions using the pair $(n,m)$. Coefficients in this expansion are read off from this by expanding the functions $\cG$ and $\cF_n$ in powers of $\psi$ and collecting terms, and for what follows what is most important is that all interactions involve non-negative powers of $c_s$ (in addition to any dependence they might also have on slow-roll parameters). Furthermore, the only interactions that arise unsuppressed by any powers of $c_s$ are those coming from the expansion of the square-root term, since all of the $\psi$-dependence in $\cF$ comes $c_s$-suppressed.

\subsection{Power-counting}
\label{sec2:powercount}

Power-counting determines how an arbitrary Feynman graph built using arbitrary interactions within the EFT depends on the ratio between the energy, $E$, of physical interest and the high-energy scales (like $M$ and $v$ above), whose small size underlies the low-energy approximation on which the EFT rests. This detailed knowledge of how different interactions and graphs contribute allows a systematic determination of which graphs are relevant to any given order in small quantities like $E/v$ and $E/M$, putting it at the heart of any effective theory's utility.

This kind of power-counting is done for correlation functions of fields in simple slow-roll models in \cite{ABSHclassic, InfPowerCount}. There it is shown which interactions and graphs contribute at any order in powers of slow-roll parameters and of small energy ratios (like $E/M$, $E/v$ and $E/M_p$, where $E \sim H$ is chosen of order the inflationary Hubble scale). The results obtained both reproduce standard lore as to the size and origins of the leading contributions, and predict which graphs and interactions provide the principal subdominant terms. The goal of this paper is to supplement these power-counting rules to keep track of any additional suppressions by powers of $c_s$. This is most easily done by expressing the Feynman rules using the variables $\psi$ and $t'$ introduced above, and counting the small factors of $c_s$ wherever they lie. The present section does so for the self-interactions within the DBI scalar sector, while those that follow also include gravitational and non-DBI interactions.

Consider therefore any interaction operator, $\cO_r$, generated in the Lagrangian \pref{eqn::effective_Lagrangian}. As this equation shows, each interaction arises partly from the expansion of $\sqrt{1-X/\Lambda^4}$ and partly through the expansion of $\cF$, and the dependence of the couplings on $c_s$ partly depends on how each of these contributes to the term of interest. For each $r$ there is a collection of pairs $(n,m)$ such that $\cO_n \cO_m$ of \pref{eqn::effective_Lagrangian} is contained within $\cO_r$. Each interaction vertex can therefore be labelled by the following non-negative numbers:
\begin{itemize}
	\item $N_{nm}$, counting the total number of lines that converge at the vertex;
	\item $N^\ssD_{n}$ number of the fields coming from the expansion of the square-root part of the action
	\item $N^\ssF_{m}$ number of fields coming from the expansion of $\cF$
	\item $d_{nm}$, counting the total number of derivatives of all types in the vertex;
	\item $\theta^\ssD_{n}$ number of {\it time} derivatives coming from the square-root part of the action
	\item $\theta^\ssF_{m}$ number of {\it time} derivatives coming from the expansion of $\cF$
\end{itemize}
The last three of these only count derivatives that act on the {\it fluctuation} field, $\phi = \Phi(\bfr, t) - \varphi(t)$, since the derivatives of the background field, $\dot \varphi$, are counted separately through the dependence on $c_s$.

These definitions imply the following relationships hold for each $n$ and $m$:
\be
  N_{nm} = N^\ssD_{n} + N^\ssF_{m}  \ge 3 \qquad \hbox{and} \qquad d_{nm} \ge \theta^\ssD_{n} + \theta^\ssF_{m} \ge 0\,,
\ee
where $\sigma_{mn} := d_{nm} - \theta^\ssD_{n} - \theta^\ssF_{m} \geq 0$ counts the total number of spatial derivatives in the vertex of interest. The inequality $N_{nm} \ge 3$ holds for all interactions because terms with $N_{nm} = 1$ cancel by virtue of the background field equations and those with $N_{nm} =2$ define the unperturbed lagrangian, relative to which the interactions are inferred. Because fields only enter into the square-root action either singly differentiated or undifferentiated, it follows that for these $\theta^\ssD_{n} \le N^\ssD_{n}$.  
   
Consider now a graph containing $\cE$ external lines, built using $V_{nm}$ vertices of type `$(n,m)$' as defined in \pref{eqn::effective_Lagrangian}. The couplings associated with these vertices contribute to this graph the following factor 
\be \label{eqn::vertex_contribution}
	\hbox{(vertices)} = \prod_{(mn)} \bigg[ M^2 v^2 \mathfrak{a}_n(c_s) \mathfrak{b}_m c_s^{N^\ssF_{m}+\theta^\ssF_{m}} \bigg( \frac{1}{v} \bigg)^{N_{nm}} \bigg( \frac{1}{M} \bigg)^{d_{nm}} \bigg]^{V_{nm}}
\ee
reflecting the fact that $\psi$ always comes with a factor of $1/v$ and that each factor of $\psi$ or $\partial_{t'}$ in $\cF$ also comes with a factor of $c_s$. The point of working with the canonically normalized variable $\psi$ is that all propagators in the graph are $c_s$-independent, and so do not complicate its counting. 

The dependence on energy scales is simplest if the graph in question involves only one important low-energy scale --- call it $q$ --- since then the dependence on $q$ of the Feynman graph can be determined (up to logarithms) using dimensional analysis\footnote{This argument is simplest when divergences are handled using dimensional regularization.} \cite{EFTreview0, EFTreview}. For instance, for simple cosmologies a correlation function typically depends on the value of the external momenta, $k$, and on the background Hubble scale, $H$, and so only one important scale arises for correlation functions evaluated with physical momenta of order\footnote{For cosmologies with $a = a(Ht)$ the conversion in DBI models from $t$ to $t' = c_s t$ implies $a = a(H' t')$ where $H' := H/c_s$ is the sound horizon, making evaluation at the sound horizon the single-scale case -- see \S\ref{sec:PCDBIgrav} for details.}  $H$. 

Following the same steps as in \cite{ABSHclassic, InfPowerCount} gives the following estimate for the dependence on $q$ of a generic amputated\footnote{For an amputated correlation function propagators for the external lines are not included, as can be appropriate for scattering amplitudes or contributions to effective couplings in an effective action.} correlation function with $\cE$ external lines 
\be \label{eqn::powercounting_dbi}
	\mathcal{A}_\cE(q) = v^2q^2 \bigg[\frac{1}{v}\bigg]^{\mathcal{E}} \bigg[ \frac{q}{4\pi v}\bigg]^{2L} \prod_{(mn)} \left[ \mathfrak{a}_n(c_s) \, \mathfrak{b}_m \,c_s^{N^\ssF_m+\theta^\ssF_m} \bigg( \frac{q}{M}\bigg)^{(d_{nm}-2)} \right]^{V_{nm}} \,,
\ee
where $L$ counts the number of loops in the graph, obtained from the definition
\be
 L = 1 + \cI - \sum_{(nm)} V_{nm} \,.
\ee 
Here, $\cI$ denotes the number of internal lines appearing in the graph and writing the power of $1/v$ in eq.~\pref{eqn::powercounting_dbi}, also uses the `conservation of ends' identity that holds for any graph:
\be
 \cE + 2\cI =\sum_{(nm)} N_{nm} V_{nm} \,.
\ee
The corresponding result for a correlation function like $\langle \phi(q_1) \cdots \phi(q_\cE) \rangle$ is obtained from the above by re-attaching propagators to the external lines ({\it i.e.}~un-amputating), giving
\be \label{eqn::powercounting_dbi_cosmology}
	\mathcal{B}_\cE(q) = \bigg(\frac{v}{q}\bigg)^2\bigg[ \frac{q}{4\pi v}\bigg]^{2L}\bigg[\frac{q^2}{v}\bigg]^{\mathcal{E}} \prod_{(mn)} \left[ \mathfrak{a}_n(c_s) \, \mathfrak{b}_m \, c_s^{N^\ssF_m+\theta^\ssF_m} \bigg( \frac{q}{M}\bigg)^{(d_{nm}-2)} \right]^{V_{nm}}\,.
\ee

In the limit $c_s \to 1$ eqs.~\pref{eqn::powercounting_dbi} and \pref{eqn::powercounting_dbi_cosmology} reproduce the power-counting results found in \cite{ABSHclassic, InfPowerCount} for generic slow-roll models. In particular, graphs built using only interactions with two or more derivatives --- {\it i.e.}~those with $d_{nm} \ge 2$ --- introduce only non-negative powers of $q/v$ and $q/M$. This is why the low-energy expansion makes sense (and ultimately what explains the simplicity of the low-energy limit). Furthermore, the small value of the combination $q/4\pi v$ is revealed as the justification for use of the semi-classical approximation; all other things being equal, zero-loop (tree) graphs dominate and each loop costs a factor of $(q/4\pi v)^2$. (In particular, classical methods are a bad approximation when $q/4\pi v$ is not small.)

The appearance of negative powers of $q/M$ for vertices with $0 \le d_{nm} < 2$ derivatives is also a well-known issue \cite{ABSHclassic, InfPowerCount} for scalar field theories, and reflects the fact that the scalar potential (and mass terms more generally) can undermine the consistency of having a light scalar appearing in the would-be low-energy limit. This does not cause problems for the low-energy limit of inflationary models because the scalar potential is usually assumed (often without explanation) not to be generic. For instance, taking all $d_{nm} = 0$ interactions to have the form 
\be \label{ShallowV}
  V(\phi) = \mu^4 \cV(\phi/v) =  \mu^4 \left[ \cV_0 + \cV_2 \, \left( \frac{\phi^2}{v^2} \right)  + \cV_4 \, \left( \frac{\phi^4}{v^4} \right) + \cdots \right]\,,
\ee
implies each $d_{nm} = 0$ vertex comes suppressed by a factor $\mu^4/(v^2 M^2)$ compared with those considered in the power-counting arguments made above, and as a result modify the $d_{nm} = 0$ contributions of \pref{eqn::powercounting_dbi} or \pref{eqn::powercounting_dbi_cosmology} to
\be \label{eqn::powercounting_dbi_potential}
	\hbox{($d_{nm} = 0$ vertices)} \propto \left[ \frac{\mu^4}{v^2M^2} \left( \frac{M}{q}\right)^{2} \right]^{V_{nm}} = \left[ \frac{\mu^4}{v^2q^2} \right]^{V_{nm}}\,,
\ee
which are small so long as $\mu^2 \lsim qv$. For many models it turns out that $v \simeq M_p$ and $q \simeq H$, and so each $d_{nm} = 0$ vertex contributes a factor of $(\mu^2/qv)^2 \simeq (\mu^2/HM_p)^2$. However, whenever the scalar potential dominates the energy density (as is usually required for successful cosmologies) the Friedmann equation implies $H^2 \simeq V/M_p^2 \simeq (\mu^2/M_p)^2$, in which case $\mu^2/HM_p \simeq \cO(1)$. For such models the inclusion of $d_{nm} = 0$ interactions becomes neutral to the validity low-energy approximation.

But for the present purposes the most important property of \pref{eqn::powercounting_dbi_cosmology} is this: it involves only non-negative powers of $c_s$. In detail this happens because of the assumption that the path integral is dominated by configurations with $\psi, \partial_{t'} \psi \sim \cO(1)$. Because of this, it can be consistent to employ semiclassical reasoning to the regime $c_s \ll 1$, at least for DBI/EFT-style actions for which the entire scalar lagrangian comes pre-multiplied by the square-root factor involving $X$. For these kinds of actions specifically, it makes sense to work to all orders in $X$ while neglecting higher derivatives like $\partial^n \Phi$ with $n \geq 2$.

\subsubsection*{Leading behaviour}

To make this all explicit, consider the special case where $\cE = 2$ and $v \simeq M \simeq M_p$ and restrict attention to the two-point correlator $\cB_2(q)$ evaluated with small external momenta $q  \ll M_p$.  In this case according to \pref{eqn::powercounting_dbi_cosmology}, the contributions least suppressed by powers of $q/M_p$, require two conditions to be satisfied: ($i$) $L = 0$  and ($ii$) $V_{nm} = 0$ for all interactions with $d_{nm} > 2$. That is to say, one works within the classical approximation using only interactions that contain two (or fewer) derivatives of the fluctuation field $\psi(\bfr, t')$. 

Inspection of \pref{DBIpsiexpn} and \pref{cFpsiexpn} gives the interactions involving at most two derivatives of the fluctuation in the DBI-EFT lagrangian, with (recalling $\Lambda^2 = vM \sim M_p^2$)
\ba
  \mL_\eff &=& - \frac{M_p^4}{c_s} \sqrt{1 - \frac{X}{M_p^4}} \; \cF \nn\\
&\supset& \left[  -v^2 M^2 + vM \sqrt{1-c_s^2} \;\psi' + \frac12 \, (\psi')^2  -  \frac12(\nabla \psi)^2 + \cdots \right] \\
&& \qquad \times \left\{ \cF_{0} + \cF_{1} \, \frac{c_s(c_s \psi)'}{Mv} + \cF_{2}\, \left[ \frac{c_s(c_s \psi)'}{Mv} \right]^2 + \cF_{3}  \, \frac{(c_s \nabla \psi)^2}{2M^2v^2} + \cdots \right\} \,,\nn
\ea
where $\cF_k = \cF_k(c_s \psi/v)$ depends only on undifferentiated fields. For small $c_s$ this is dominated by the terms independent of $c_s$, which are
\be
  \mL_\eff^{(0)} = \left[  -v^2 M^2 + vM \,\psi' + \frac12 \, (\psi')^2  -  \frac12(\nabla \psi)^2 \right]  \cF_{00}  \,,
\ee
where $\cF_{00}$ denotes the $\psi$-independent part of $\cF_0$ (since any explicit power of $\psi$ comes pre-multiplied by $c_s$). The only contributions not suppressed by powers of $c_s$ come from expanding the square-root action (so two-derivative terms coming from expanding $\cF$ are all subdominant in $c_s$). The leading contribution in the $q/M_p$ and $c_s$ expansion is therefore obtained using classical calculations made with the original DBI action (as is indeed common practice in the literature).

Eq.~\pref{eqn::powercounting_dbi_cosmology} gives the size of the leading contribution to $\cB_2$ in powers of $q/M_p$ to be
\be \label{eqn::powercounting_dbi_cosmologyE2}
	\mathcal{B}_2(q) \simeq q^2   \prod_{{(mn) \atop d_{nm} = 0}} \left[ \mathfrak{a}_n(c_s) \, \mathfrak{b}_m \, c_s^{N^\ssF_m} \, \frac{\mu^4}{M_p^2 \,q^2}  \right]^{V_{nm}} \prod_{{(mn) \atop d_{nm} = 2}} \left[ \mathfrak{a}_n(c_s) \, \mathfrak{b}_m \, c_s^{N^\ssF_m+\theta^\ssF_m} \right]^{V_{nm}}\,.
\ee
For inflationary applications one would use a scalar potential given by \pref{ShallowV} with known order-unity coefficients $\cV_k$. The inflationary assumption that this potential dominates the gravitational response implies the Hubble scale is $H \sim \mu^2/M_p$, making the factor $\mu^4/(M_p^2 \, q^2) \sim H^2/q^2$ (and so order unity if $q \sim H$).\footnote{The factor $\mu^4/(M_p^2q^2)$ is instead order $c_s^2$ if evaluated at the sound horizon, for which $q \sim H/c_s$.}

The first subdominant term, suppressed by a single power of $c_s$, arises if $\cF_0 =  \cF_{00} + \cF_{01} \, (\phi/v) + \cdots$ contains a linear term in $\phi$, and is given by  
\be
  \mL_\eff^{(1)} =  \left[  -v^2 M^2 + vM \,\psi' + \frac12 \, (\psi')^2  -  \frac12(\nabla \psi)^2 \right]  \cF_{01} \, \frac{c_s\psi}{v} \,, 
\ee
and so on, to any order in $H/M$, $H/v$ and $c_s$. Any dependence on slow roll parameters can be treated similarly, along the lines of \cite{ABSHclassic}.

\section{Power-counting (including gravity)}
\label{sec:PCDBIgrav}

The previous section shows how the square-root factor of the DBI-EFT action allows a sensible perturbative expansion despite working to all orders in $X$ in the square root provided that the sound speed is small: $c_s \ll 1$. This section extends this result to include terms in $\mathfrak{L}_\eff$ not multiplied by the square root factor, including in particular the Einstein-Hilbert action, describing the metric fluctuations. 

To see how this works, consider the following generalization of \pref{L0def} to an effective action coupling $\Phi$ to the metric,
\be \label{L0defgrav}
 \mathfrak{L}_{\rm eff} = - \sqrt{-g} \left[ \sqrt{1 - \frac{X}{v^2M^2}} \;\cF(\Phi, \partial \Phi, \partial^2 \Phi, \cR , \dots) 
  +  \cK(\Phi, \partial \Phi, \partial^2 \Phi, \cR , \dots) \right] \,,
\ee
where $\cR = g^{\mu\nu} \cR_{\mu\nu}$ and $X = - g^{\mu\nu} \partial_\mu \Phi \, \partial_\nu \Phi$ while $\cF$ and $\cK$ are arbitrary functions to be treated in a standard derivative expansion, including both derivatives of the scalar $\Phi$ as well as powers of the metric's Riemann tensor, ${\cR^\mu}_{\nu\lambda\rho}$, and its contractions and derivatives. Among the lowest-dimension terms in $\cK$ is the Einstein-Hilbert action
\be \label{L0defgrav1}
  \cK = \frac{M_p^2}2 \, \cR + \cdots \,.
\ee
In writing \pref{L0defgrav1} the metric is assumed to be Weyl-rescaled to Einstein frame, defined by the absence of any $\Phi$-dependence in the Einstein-Hilbert action. A field redefinition is also used to eliminate any $\Phi$-dependence in front of $X$ within the square root.  

\subsection{New complications}

Coupling gravity to the DBI scalar introduces two new issues that can complicate the power-counting story. First, some terms in the action --- those involving $\cK$ in \pref{L0defgrav} --- do not come with the overall square-root factor. Second, the fact that the graviton and scalar propagate with different speeds makes the counting of the $c_s$-dependence of a graph more subtle. 

Having terms in the action not proportional to the square-root factor is not that serious a complication, provided that they are treated using a standard derivative expansion ({\it i.e.}~not treated to all orders in $X$). By contrast, having different propagation speeds for scalar and metric tensor fluctuations is an important complication because it becomes impossible to simultaneously scale $c_s$ out of the quadratic part of the fluctuation lagrangian for {\it both} the scalar and the metric. This cannot be done because the free scalar action, $\dot \phi^2 - c_s^2 (\nabla \phi)^2$, depends differently on $c_s$ than does the free part of the Einstein-Hilbert action, $\dot h^2 - (\nabla h)^2$ for the metric tensor fluctuation $h$. This inability to scale $c_s$ completely out of all propagators forces a change in the power-counting for the perturbative expansion in $c_s$, making contributions with gravity loops less suppressed by powers of $c_s$ than they would otherwise be. 

To see more explicitly why this might happen, expand the above effective action about a homogeneous classical solution,
\be \label{eqn::expansion_around_classical}
	g_{\mu\nu} = \bar{g}_{\mu\nu} + \frac{2h_{\mu\nu}}{M_{p}} \quad \hbox{and} \quad
	\Phi = \varphi(t) + \phi = \varphi(t'/c_s) + c_s \psi \,,
\ee 
%
where $\bar{g}_{\mu\nu} \, \exd x^\mu \exd x^\nu = - \exd t^2 + a^2(t) \, \exd \sigma^2$ is a spatially flat FRW metric, with scale factor $a(t)$ obtained together with $\varphi(t)$ by solving the Einstein-scalar background field equations. As before, the dominant term in the free action for the scalar fluctuation comes by expanding the square-root term, from which all dependence on $c_s$ is removed by rescaling $\phi \to \psi = \phi/c_s$ and $t' = c_s t$, leading to a quadratic scalar action of the form 
\be
  S_\psi^{(2)} = \int \exd t \exd^3 x \, a^3(t) \frac1{2c_s^3} \left[ \dot\phi^2 - \frac{c_s^2}{a^2(t)} (\nabla \phi)^2 \right]  = \frac12 \int \exd t' \exd^3 x \, a^3(t'/c_s) \left[ (\psi')^2 - \frac{1}{a^2(t'/c_s)} (\nabla \psi)^2 \right] \,,
\ee
which for simplicity drops terms involving $\exd c_s/\exd t$, since for inflationary applications these are usually slow-roll suppressed (and can be included as required without difficulty).

Because of the rescaling of time this introduces factors of $c_s$ into the quadratic metric fluctuations, which for tensor modes, $h_{ij}$, schematically behaves like
\be
  S_\ssE^{(2)} = \frac12 \int \exd^4x \; a^3(t) \left[ \dot E^2 - \frac{1}{a^2(t)} (\nabla E)^2 \right] = \frac12 \int \frac{\exd t' \exd^3x}{c_s } \; a^3(t'/c_s) \left[ c_s^2 (E')^2 - \frac{1}{a^2(t'/c_s)}(\nabla E)^2 \right] \,,
\ee
where $h_{ij} = a^2(t) E_{ij}$. These expressions show that rescaling $h_{ij}$ (or $E_{ij}$) can at best remove $c_s$ from the $(E')^2$ or the $(\nabla E)^2$ term, but not both. Rescaling $E \to \hat E := E/\sqrt{c_s}$ then gives
\be \label{S0hhat}
  S_\ssE^{(2)}  = \frac12 \int \exd t' \exd^3x \; a^3(t'/c_s)  \left[ c_s^2 (\hat E')^2 - \frac{1}{a^2(t'/c_s)}(\nabla \hat E)^2 \right] \,,
\ee
which suggests that the term $c_s^2 (\hat E')^2$ can be treated as perturbatively small in the $c_s \ll 1$ limit, and so be included amongst the interactions rather than the unperturbed part of the action. Dropping the time derivative in this way reflects how the influence of graviton fluctuations effectively becomes instantaneous on time-scales relevant to the much slower motion of the $\psi$ field.  

It is tempting to conclude that an expansion in powers of $c_s$ should also work here, provided that all interactions also involve no negative powers of $c_s$ once expressed using the variables $\psi$, $\hat E$ and $t'$. Indeed this conclusion would be correct if it were true that the path integral were always dominated by configurations for which both $(\nabla \hat E)^2$ and $(\hat E')^2$ were similar in size (order unity), and not themselves enhanced as $c_s \to 0$. But because there is a functional integral over $\hat E$ there is no guarantee that the integral's saddle point need be dominated in this regime. Furthermore, in this regime the unperturbed action does not depend on $(\hat E')^2$ at all (because of the suppression by $c_s^2$), so $e^{iS_h^0}$ cannot suppress rapidly varying modes from contributing to and ultimately dominating the path integral. 

An indication that problems might arise at high frequency can be found by keeping both the time and space derivatives of \pref{S0hhat} in the unperturbed action when evaluating Feynman graphs, since if the small-$c_s$ limit is nonsingular this alternative way of computing should capture the same physics as when $c_s^2 (\hat E')^2$ is treated as a perturbation. For instance, evaluating a loop graph involving both scalar and tensor propagators on a flat-space background can give momentum-space loop integrations of the schematic form
\be
 \int \frac{\exd^4 k}{(2\pi)^4} \; \frac{i}{ k_0^2 - k^2 + i \epsilon} \; \frac{i}{ c_s^2 k_0^2 - k^2 + i \epsilon} \; N(k_0, k)
\ee
where $k_\mu = \{ k_0, \bfk \}$ is a loop momentum to be integrated when evaluating the graph and both a scalar and tensor propagator are displayed explicitly, with $k = |\bfk|$. The factor $N(k_0, k)$ represents the contribution of the rest of the graph, with external momenta neglected here for simplicity. 

The dangerous regions of integration for small $c_s$ are then displayed by evaluating the $k_0$ integral by residues, which involves replacing $k_0$ everywhere by either $k_0 = k$ or $k_0 = k/c_s$, leading to factors of the form:
\be \label{PoleProb}
 \int \frac{\exd^3 k}{(2\pi)^3} \;  \frac{1}{ 2k}  \left[ \frac{N(k,k)}{ (c_s^2 -1) k^2} +   \frac{N(k/c_s,k)}{ (c_s^{-2} -1) k^2} \right] \,.
\ee
For the metric the regime $k_0 \simeq k/c_s$ explicitly violates the assumption made above that $c_s^2 (\hat E')^2$ can be dropped relative to $(\nabla \hat E)^2$. Contributions from this part of the integral therefore can give enhancement by powers of $1/c_s$ relative to a naive inspection of the interactions, depending on how $N(k_0, k)$ depends\footnote{For instance, no enhancement at small $c_s$ occurs if $N$ contains a factor of $c_s$ for every factor of $k_0$, although this is something we know is not true if $N$ is built using interactions coming from the square-root part of the scalar action (which can involve $k_0$ without also a $c_s$, even once written in terms of the new fields).}  on $k_0$.  (A similar interplay between slowly moving and relativistic quanta also complicates power-counting for QED applied to systems -- like atoms -- containing slowly-moving electric charges \cite{NRQED, Labelle:1996en, Luke:1996hj, Grinstein:1997gv, pNRQED}.)

The potential appearance of factors of $1/c_s$ in $N(k/c_s,k)$ need not mean that the series expansion in $c_s$ completely breaks down. This is because any dangerous graph must involve both the gravitational and the DBI sector, since the problematic poles involve graviton loops but the interactions unsuppressed by powers of $c_s$ arise in the DBI scalar sector. In any particular graph one must connect the graviton sector to the scalar DBI sector and the small-$c_s$ limit can be sensible if this connection involves sufficiently many powers of $c_s$. As is explored in more detail in \S\ref{dangerousloop}, this is what we believe happens for DBI inflation.

\subsection{Scalar-metric mixing}

The previous section argues that counting powers of $c_s$ can be more complicated in graphs that involve both DBI scalars and the metric's propagating tensor modes. We next ask whether similar things can happen when we ask how $\psi$ interacts with scalar metric fluctuations. We do so within Newtonian gauge, for which the metric (including fluctuations) is given by
\begin{equation}\label{eq: merticNewtonian}
    \exd s^{2}= \left(-1-\frac{2A}{M_{p}} \right) \exd t^{2} + a^2(t) \left[ \left( 1-\frac{2B}{M_{p}}  \right) \delta_{ij}+ \frac{2E_{ij} }{M_p} \right] \, \exd x^{i}\exd x^{j},
\end{equation}
where the metric's scalar perturbations $A$ and $B$, respectively, are the Newtonian and relativistic potentials, and the traceless fluctuation $E_{ij}$ contains the transverse gravitational waves. When $E_{ij} = 0$ the measure becomes $\sqrt{-g} = a^3 \left( 1 - {2B}/{M_p} \right)^{3/2} \sqrt{ 1 + {2A}/{M_p}}$.

To check the $c_s$ expansion we follow the steps of the previous sections: first scale as much of the $c_s$-dependence as possible from the quadratic part of the fluctuation action, and then identify the $c_s$-dependence of the interactions. Expanding the Einstein-Hilbert action to quadratic order in $A$ and $B$ in this gauge leads to the following quadratic `free' action for the scalar part of the metric fluctuations, 
\ba  \label{eq: EHquadraticPart}
       S_{EH}^{(2)} &=& \frac{1}{2}  \int \exd t \exd^{3}x \, a^{3} \left[ \frac{2 \nabla B \cdot (2 \nabla A-\nabla B)}{a^2} - 12H \dot B  (A+B) +   9H^2 (A+B)^2  -6 \dot B^{2}  \right] \nn\\
       &=& \frac{1}{2}  \int \exd t' \exd^{3}x \, a^{3} \left[ \frac{2 \nabla B \cdot (2 \nabla A-\nabla B)}{a^{2} c_{s}}- 12 {B'} H (A+B)+    \frac{9H^2}{c_{s}} (A+B)^2  -6  c_{s} ({B'})^{2} \right] \,
\ea
where the second line changes variables $t \to t' = c_s t$ and primes denote $\exd / \exd t'$. $H = \dot a/a$ denotes the background derivative, $H$, with respect to $t$ (just as was also done for $\dot\varphi$ when keeping the background factors of $c_s$). For later use we also note that the $t'$-dependence of the background geometry is characterized by the inverse sound horizon, defined by $d_s = H_s^{-1}$ with
\be \label{SoundHorizon}
  H_s := \frac{1}{a}\left( \frac{\exd a}{\exd t'} \right) = \frac{H}{c_s} \gg H\,.
\ee

To this must be added the quadratic part of the DBI action, including the metric fluctuations. The first step in deriving this is to insert \eqref{eq: merticNewtonian} into the square-root part of the action, using
\ba \label{eq: korak10}
    1 - \frac{X}{\Lambda^4} &=&1 - \frac{1}{\Lambda^4} \left[ \frac{(\dot \varphi + \dot \phi)^2}{1+2A/M_p} - \frac{1}{a^2} \left( \frac{ \nabla \phi \cdot \nabla \phi}{1 - 2 B/M_p} \right)\right] \\
    &=&1 -  \left[ \frac{1- c_s^2 + 2\sqrt{1 - c_s^2}\; (\dot \phi/\Lambda^2) + (\dot\phi^2/\Lambda^4)}{1+2A/M_p^2} - \frac{1}{a^2\Lambda^4} \left( \frac{ \nabla \phi \cdot \nabla \phi}{1 - 2 B/M_p^2} \right) \right]\nn\\
    &=&  \frac{ c_s^2 +2(A/M_p^2) -2c_s^2\sqrt{1 - c_s^2}\; (\psi'/\Lambda^2) - c_s^4 [(\psi')^2/\Lambda^4]}{1+2A/M_p^2} + \frac{c_s^2}{a^2\Lambda^4} \left( \frac{ \nabla \psi \cdot \nabla \psi}{1 - 2 B/M_p^2} \right)  \,,\nn
\ea
and so
\ba \label{eq: korak11}
    S_\DBI  &=&  - \Lambda^4 \int \exd t \exd^3x \; \sqrt{-g} \sqrt{1 - \frac{X}{\Lambda^4}} \nn\\
    &=&  - \Lambda^4 \int \exd t' \exd^3x \;  a^3 \left(1 -\frac{ 2B}{M_p} \right) \left\{\left( 1 + \frac{2A}{M_p c_s^2} -2\sqrt{1 - c_s^2}\; \frac{\psi'}{\Lambda^2} \right. \right. \\ 
    && \qquad\qquad\qquad\qquad \left. \left. - \frac{c_s^2 (\psi')^2}{\Lambda^4} \right) \left(1 -\frac{ 2B}{M_p} \right) + \frac{\nabla \psi \cdot \nabla \psi}{a^2\Lambda^4} \left( 1 +\frac{2 A}{M_p} \right) \right\}^{1/2} \,.\nn
\ea
Expanding $S_\DBI$ to second order then gives the DBI contribution to the quadratic action for fluctuations:
\be \label{eq: DBIquadraticPart}
    S_{DBI}^{(2)} =  \int \exd t' \exd^{3}x \; a^{3}
    \left[ \frac{(\psi')^2}2 -  \frac{(\nabla \psi)^2}{2a^2} 
   -\frac{ \Lambda^{2}}{M_p} \sqrt{1-c_{s}^{2}} \left(\frac{A}{c_{s}^{2}} + 3B\right) {\psi'} 
    + \frac{\Lambda^{4}}{M^{2}_{p}}  \left(\frac{A^2}{2c_s^4} + \frac{3AB}{c_{s}^{2}} - \frac{3B^2}2 \right)
    \right] \,.
\ee

Inspection of the Einstein-Hilbert action \pref{eq: EHquadraticPart} suggests defining the new variable $\hat B := B/\sqrt{c_s}$ while the DBI action \pref{eq: DBIquadraticPart} suggests defining $\hat A := A/c_s^2$. With these variables the most singular terms of $S^{(2)} = S^{(2)}_\EH + S^{(2)}_\DBI$ in the limit $c_s \to 0$ are $\cO(1)$, with
\ba \label{eq:QuadraticAction}
    S^{(2)} &=&  \int \exd t' \exd^{3}x \; a^{3}
    \left[ \frac{(\psi')^2}2 -  \frac{(\nabla \psi)^2}{2a^2} 
   - \frac{Mv}{M_p} \sqrt{1-c_{s}^{2}} \left(\hat{A}+ 3\sqrt{c_s} \hat B\right) {\psi'} \right. \nn\\
   && \qquad\qquad + \frac{M^2v^2}{M^{2}_{p}}  \left(\frac{\hat A^2}2 + {3 \sqrt{c_s} \hat A \hat B} - \frac{3c_s}2 \hat B^2 \right) +  \frac{ \nabla \hat B \cdot (2 c_s^{3/2} \nabla \hat A-\nabla \hat B)}{a^{2}}\\
    && \qquad\qquad\qquad\qquad  \left. - 6c_s {\hat B'} H (c_s^{3/2} \hat A+ \hat B)+  \frac{9H^2}2 (c_s^{3/2} \hat A+ \hat B)^2  -3  c_{s}^2 ({\hat B'})^{2}   \right] \nn\\
   &\simeq&  \int \exd t' \exd^{3}x \; a^{3}  \left[\frac{ (\psi')^2}2 -  \frac{(\nabla \psi)^2}{2a^2} 
   - \frac{Mv}{M_p}  \, \hat{A} {\psi'}  + \frac{M^2v^2}{M_p^2}  \, \frac{\hat{A}^2 }2   - \frac{ \nabla \hat B \cdot  \nabla \hat B}{a^{2}} +  \frac{9H^2}2  \hat B^2     + \cO(\sqrt{c_s}) \right]\,,\nn
\ea
which assumes for simplicity that $\Lambda^2 = Mv$. The final approximate equality keeps only terms\footnote{The term involving $H^2 \hat B^2 = c_s^2 H^2_s \hat B^2$ can also be dropped if $c_s \to 0$ while fixing $H_s$ rather than $H$.} unsuppressed by positive powers of $c_s$. Eq.~\pref{eq:QuadraticAction} depends only on spatial derivatives of $\hat B$ in the limit that $c_s \to 0$, similar to what was found above for tensor fluctuations. 

\subsection{Power-counting}
\label{PowercountingDBIGR}

Written in terms of the variables $t'$, $\psi$, $\hat A$ and $\hat B$ all terms in the DBI-EFT action involve only non-negative powers of $c_s$. This makes it seem plausible that physical quantities should allow sensible expansions in powers of $c_s$, in addition to the usual expansions in powers of $H/M_p$ and slow-roll parameters. The main question then is whether the saddle points for the path integral actually occur for values of $\psi$, $\hat A$ and $\hat B$ that are order unity in the small-$c_s$ limit --- along the lines discussed around eq.~\pref{PoleProb} above.

We now repeat the exercise of counting the factors of $c_s$ and the low-energy scale $q$ that appear in specific Feynman graphs, focussing on the leading powers since our main interest is in whether or not a series in low energies and in $c_s$ exists at all. More detailed formulae that work to any specific order in $c_s$ can be found in precisely the same way, though require somewhat more effort. Since we use $t'$ as our time variable, the natural scale of the background is $H_s = a'/a$ since this appears in the propagator for fields like $\psi(t')$ in the same way as does $H$ for a canonical field $\phi(t)$. We therefore assume all external scales are of order $q \sim H_s = H/c_s$ when doing our dimensional analysis. Furthermore, for simplicity we take $v \sim M_{p}$ with the scalar potential chosen to dominate the gravitational response (as is usually the case in inflationary models) so that the scales in the potential satisfy $\mu^2 \sim H M_p \sim H_s M_p/c_s $. This last assumption allows us to avoid the problems with zero-derivative interactions discussed around eq.~\pref{ShallowV}. 

To start, we treat all terms multiplied by positive powers of $c_{s}$ as interactions and return to the appearance of factors of $1/c_s$ due to poles in gravity loops in the next section. Because the quadratic action in the last line of \pref{eq:QuadraticAction} has an unorthodox form once powers of $c_s$ are dropped, we make the power-counting argument directly in position space rather than going to momentum space. The fact that $\hat A$ and $\psi$ mix in the quadratic action does not matter for power-counting purposes, apart from the recognition that $\hat A$ always enters pre-multiplied by a factor of $Mv/M_p$, and so this factor must be kept track of when evaluating at the path integral's saddle point. This can be done either explicitly in the Feynman rules for the propagator or by rescaling to $\tilde A = (Mv/M_p) \hat A$. Or, because $v \sim M_p$, $\tilde A = M \hat A$.

{} From here on the counting proceeds much as for the scalar-only DBI model described above, so we highlight mostly the parts that change. Just as for the scalar-only case interactions from the DBI part of the action arise as a product of interactions coming from expanding the square root and from expanding the function $\cF$:
\ba \label{DBIcoeffdef}
S_{DBI}  &=& M^{2} M_p^{2}\sum_{m}
            \mathfrak{a}_{m}(c_{s}) \cO_{m} \left(\frac{\hat A}{M_{p}},\frac{\sqrt{c_s}\, \hat B}{M_{p}}
        , \frac{\psi'}{MM_p} , \frac{\nabla \psi}{MM_p} ,\cdots\right) \nn\\
        &&\qquad\qquad\qquad\qquad\qquad\qquad \times \sum_{n} \mathfrak{b}_{n}   \, \cO_{n}\left(\frac{c_s^2 \hat A}{M_{p}},\frac{\sqrt{c_s} \hat B}{M_{p}},\frac{c_s \psi}{M_p}, \frac{c_s\partial_{t'}}{M}, \frac{\nabla}{M}, \cdots\right) \nn\\
        &=& M^{2}M_p^{2}\sum_{m}
            \mathfrak{a}_{m}(c_{s}) \cO_{m} \left(\frac{\tilde A}{MM_p},\frac{\sqrt{c_s}\, \hat B}{M_{p}}
        , \frac{\psi'}{MM_p} , \frac{\nabla \psi}{MM_p} ,\cdots\right) \\
        &&\qquad\qquad\qquad\qquad\qquad\qquad \times \sum_{n} \mathfrak{b}_{n}   \, \cO_{n}\left(\frac{c_s^2 \tilde A}{MM_p},\frac{\sqrt{c_s} \hat B}{M_{p}},\frac{c_s \psi}{M_p}, \frac{c_s\partial_{t'}}{M}, \frac{\nabla}{M} \right)\,,\nn
\ea
where the operators $\cO_{n}$ and $\cO_m$ schematically represent a basis of interaction operators respectively coming from the square-root term and the function $\cF$ of the DBI Lagrangian. The second equality of \pref{DBIcoeffdef} comes from rescaling $\hat A \to \tilde A$ so that the quadratic action $S^{(2)}$ does not depend on the UV scales $M$ and $M_p$ (see the last line of eq.~\pref{eq:QuadraticAction}). Although not written explicitly in \pref{DBIcoeffdef}, a similar $c_s$ dependence arises in $\mL_\DBI$ and $\mL_\EH$ for precisely the same reasons. 

The coefficients $\mathfrak{a}_{m}(c_s)$ and $\mathfrak{b}_{n}(c_{s})$ are dimensionless and calculable as in earlier sections in the pure-scalar case. Notice that -- unlike $\psi$ and $\hat A$ -- a factor of $\sqrt{c_s}$ remains with $\hat B$ in the square-root action. This is because $\hat B$ acquires its normalization from the Einstein-Hilbert part of the action rather than the square root part, and so there is no absorption of the factors of $c_s$ by other rescalings. 

A similar expansion (without the square root) also comes from expanding the function $\cK$ in powers of fluctuations, which we collectively call the `Einstein-Hilbert' interactions even though they also contain all interactions not accompanied by a square-root factor, including higher curvatures and functions like $W(\Phi)$ in the original DBI model:
\ba \label{EHcoeffdef}
S_\EH  &=& \frac{M^{2} M_p^2}{c_s} \int \exd^3x\, \exd t' \sum_{r}
            \mathfrak{c}_{r} \cO_{r} \left(\frac{c_s^2 \hat A}{M_{p}},\frac{\sqrt{c_s} \hat B}{M_{p}},\frac{c_s \psi}{M_p}, \frac{c_s\partial_{t'}}{M}, \frac{\nabla}{M} \right) \nn \\
          &=& \frac{M^{2}M_p^2}{c_s} \int \exd^3x\, \exd t' \sum_{r}
            \mathfrak{c}_{r} \cO_{r} \left(\frac{c_s^2 \tilde A}{MM_p},\frac{\sqrt{c_s} \hat B}{M_{p}},\frac{c_s \psi}{M_p}, \frac{c_s\partial_{t'}}{M}, \frac{\nabla}{M} \right)   \,.
\ea
The overall factor of $1/c_s$ here comes from the rescaling $\exd t = \exd t'/c_s$, which is not absorbed here into the square-root term (as it is for $\cF$). The dependence on tensor modes, $E_{ij}/M_p = \sqrt{c_s} \, \hat E_{ij}/M_p$, is schematically similar to the dependence on $B$, so far as their scaling with $c_s$ and $M_p$ is concerned.\footnote{Because $S_\EH$ is normalized by $M^2 M_p^2$ it overestimates the size of higher-curvature terms, which are typically suppressed only by powers of $M < M_p$. As a result the dimensionless coefficient should be regarded as being $\mathfrak{c}_r \propto M^2/M_p^2$ for vertices involving more than two derivatives (see \cite{GREFT, InfPowerCount, ABSHclassic} for details).}

Consider next the contributions to a correlation function, $\cB_{\cE_\ssA, \cE_\ssB, \cE_\phi}$, ({\it i.e.}~ an unamputated graph) involving $\cE_\ssA$ external $A$ lines, $\cE_\ssB$ external $B$ lines and $\cE_\phi$ external $\phi$ lines. Further, suppose the graph involves $L$ loops and that it contains $V_{nm}$ of the vertices labelled by $(nm)$ coming from \pref{DBIcoeffdef}, as well as by $V_r$ vertices obtained from \pref{EHcoeffdef}. Inspection of the Feynman rules shows that vertices of the graph that come from $S_\DBI$ contribute the following powers of $M$, $M_p$, $v$ and $c_s$,
\ba
  \hbox{DBI} \sim \prod_{(mn)} 
    \left[\mathfrak{a}_m(c_s) \mathfrak{b}_n M^2 M_p^2 \, c_s^{\theta^\cF_{n}}\left( \frac{H_s}{M} \right)^{d_{mn}} c_s^{N^{\psi\cF}_{n}+N^{B}_{nm}/2} \left( \frac{1}{M_p} \right)^{N^\psi_{mn}+ N^B_{mn}} c_s^{2N^{A\cF}_{n}}  \left( \frac{1}{MM_p} \right)^{N^A_{mn}} \right]^{V_{mn}}  \,,
\ea
where the product is over all of the interactions that contribute vertices somewhere in the graph. The analogous factors that come from $S_\EH$ are
\ba
   \hbox{EH} \sim \prod_r \left[\mathfrak{c}_r  \frac{M^2M_p^2}{c_s} \, c_s^{\theta_r} \left( \frac{H_s}{M} \right)^{d_{r}} \left( \frac{c_s}{M_p} \right)^{N^\psi_{r}}  \left( \frac{\sqrt{c_s}}{M_p} \right)^{N^B_{r}}  \left( \frac{c_s^2}{MM_p} \right)^{N^A_{r}} \right]^{V_r} \,.
\ea
These expressions use the following definitions for the number of derivatives and fields:
\begin{itemize}
  \item $d_{mn}$ is the total number of derivatives coming from the DBI part of the action.;
  \item $d_{r}$ is the total number of derivatives coming from an `Einstein-Hilbert' (EH) vertex;
  \item $\theta^{\cF}_{n}$ is the number of time derivatives coming from $\cF$ in the DBI vertex;
  \item $\theta_{r}$ is the total number of time derivatives coming from the EH vertex;
  \item $N^{A}_{mn}$ and $N^{\psi}_{mn}$ are respectively the total number of the fields $A$ or $\psi$ that appear in the Feynman graph, within one of the $(m,n)$ vertices. 
  \item $N^{A}_{r}$ and $N^{\psi}_{r}$ are the total number of the fields $A$ or $\psi$ appearing in the graph that come from an EH vertex. 
  \item $N^{A\cF}_{n}$ and $N^{\psi \cF}_{n}$ are the number of the fields $A$ and $\psi$ coming from the factor $\cF$ in the DBI vertex. 
    \item $N^{B}_{mn}$ and $N^{B}_{r}$ count the total number of the fields $B$ or $E_{ij}$, appearing in all vertices either in the DBI or EH lagrangians.
\end{itemize}

Because the fields are normalized so that only low-energy scales appear in propagators, the rest of the dimensions of the graph must be an appropriate power of the low-energy scale $q \sim H_s$ that restores the proper dimensions. This leads to the following expression for a correlation function involving $\cE_A$ external $A$ lines, $\cE_B$ external $B$ lines and $\cE_\phi$ external $\phi$ lines -- which has dimension (mass)${}^{\cE_\phi +\cE_A + \cE_B}$:
\ba 
 \cB_{\cE_\phi \cE_A \cE_B}(H_s) &\sim& \left\langle \phi^{\cE_\phi} A^{\cE_A} B^{\cE_B} \right\rangle \sim c_s^{\cE_\phi + 2\cE_A + \cE_B/2} \left( \frac{1}{M} \right)^{\cE_A} \left\langle \psi^{\cE_\phi} \tilde A^{\cE_A} \hat B^{\cE_B} \right\rangle \nn\\
 &\sim& c_s^{\cE_\phi + 2\cE_A + \cE_B/2} H_s^{\cE_\phi + \cE_A + \cE_B} \left( \frac{H_s}{M} \right)^{\cE_A}  \prod_{(mn)}   \left[\mathfrak{a}_m(c_s) \mathfrak{b}_n \left( \frac{M^2 M_p^2}{H_s^4} \right)\, c_s^{\theta^\cF_{n}}\left( \frac{H_s}{M} \right)^{d_{mn}} \right. \nn\\
 && \qquad   \left. \times \, c_s^{N^{\psi\cF}_{n}+N^{B}_{mn}/2} \left( \frac{H_s}{M_p} \right)^{N^\psi_{mn}+ N^B_{mn}} c_s^{2N^{A\cF}_{n}}  \left( \frac{H_s^2}{MM_p} \right)^{N^A_{mn}} \right]^{V_{mn}}  \\
 && \qquad \qquad \times  \prod_r \left[\mathfrak{c}_r \left( \frac{M^2M_p^2}{c_sH_s^4}\right) \, c_s^{\theta_r} \left( \frac{H_s}{M} \right)^{d_{r}} \left( \frac{c_sH_s}{M_p} \right)^{N^\psi_{r}}  \left( \frac{\sqrt{c_s}\, H_s}{M_p} \right)^{N^B_{r}}  \left( \frac{c_s^2H_s^2}{MM_p} \right)^{N^A_{r}} \right]^{V_r} \,.\nn
\ea

Grouping terms gives the total dependence on $H_s/M_p$ as $(H_s/M_p)^P$ with
\ba
  P &=& \sum_{mn} (N^\phi_{mn} + N^A_{mn} + N^B_{mn} - 2)V_{mn} + \sum_r (N^\phi_r + N^A_r + N^B_r - 2) V_r  \nn\\
  &=& \cE + 2 \cI - 2\sum_{mn} V_{mn} - 2\sum_r V_r = \cE + 2L - 2 \,,
\ea
which uses the `conservation of ends' identity (that holds for an arbitrary graph) for lines of type $\psi$, $A$ and $B$:
\be
  \cE^i + 2 \cI^i = \sum_{mn} N_{mn}^i V_{mn} + \sum_r N_r^i V_r \quad \hbox{for $i = \phi, A, B$} \,,
\ee
where $\cE = \cE^\phi + \cE^A + \cE^B$ is the total number of external lines, $\cI = \cI^\phi + \cI^A + \cI^B$ is total number of internal lines and $L$ is the definition of the number of loops in the graph:
\be
 L = 1 + \cI - \sum_{mn} V_{mn} - \sum_r V_r \,.
\ee
Similarly, the total power of $H_s/M$ is $(H_s/M)^Q$ with
\ba
  Q &=& \cE^A + \sum_{mn} (N^A_{mn} + d_{mn}  - 2)V_{mn} + \sum_r (N^A_r  + d_r - 2) V_r  \nn\\
  &=& 2\cE^A + 2 \cI^A + \sum_{mn} (d_{mn} - 2) V_{mn} + \sum_r (d_r -2) V_r  \,.
\ea
The overall power of $c_s$ is given by $c_s^R$ with
\ba
 R &=& \cE^\phi + 2\cE^A + \frac{\cE^B}2 + \sum_{mn} \left(\theta^\cF_n + N^{\phi\cF}_n + 2N^{A\cF}_n + \frac{N^{B}_{mn}}2 \right) V_{mn} \nn\\
 && \qquad\qquad + \sum_{r} \left(\theta_r + N^{\phi}_r + 2N^{A}_r + \frac{N^{B}_r}2 - 1 \right) V_{r} \\
 &=&\cE^\phi + 2\cE^A + \cE^B + \cI^B + \sum_{mn} \left(\theta^\cF_n + N^{\phi\cF}_n + 2N^{A\cF}_n \right) V_{mn} \nn\\
 && \qquad\qquad + \sum_{r} \left(\theta_r + N^{\phi}_r + 2N^{A}_r  - 1 \right) V_{r}  \,.\nn
\ea

Combining everything leads to the power-counting estimate for correlation functions in the DBI EFT:
\ba \label{MasterBresult}
 \cB_{\cE_\phi \cE_A \cE_B}(H_s) 
 &\sim& \frac{M_p^2}{H_s^2} \,  \left( \frac{H_s}{4\pi M_p} \right)^{2L } \left( \frac{c_s H_s^2}{M_p} \right)^{\cE^\phi}  \left( \frac{c_s^2H_s^3}{M M_p} \right)^{\cE^A}  \left( \frac{\sqrt{c_s} \, H_s^2}{M_p} \right)^{\cE^B}  \nn \\
 && \qquad \qquad \times  \prod_{(mn)}   \left[\mathfrak{a}_m(c_s) \mathfrak{b}_n \left( \frac{H_s}{M} \right)^{N^A_{mn} + d_{mn} - 2}  c_s^{\theta^\cF_n + N^{\phi\cF}_n + 2 N^{A\cF}_n + N^{B}_{mn}/2} \right]^{V_{mn}}  \\
 && \qquad\qquad\qquad\qquad \times \prod_r \left[\mathfrak{c}_r   \left( \frac{H_s}{M} \right)^{N^A_r + d_{r} - 2}  c_s^{\theta_r + N^{\phi}_r + 2 N^{A}_r + N^{B}_r/2 - 1} \right]^{V_r} \,.\nn
\ea

Two sources of $c_s$-dependence are not explicit in this expression. One comes from graviton loops, since these in general introduce additional factors of $1/c_s$ due to contributions from energy poles where the graviton is on shell, as discussed more explicitly below. The other comes from the contribution of zero-derivative interactions, since for scalar potentials of the form of \pref{ShallowV} each vertex contributes factors of the form $(H/H_s)^2 = c_s^2$. To see this write the contribution of zero-derivative vertices explicitly --- see \pref{eqn::powercounting_dbi_potential} --- to get:
\be
    \prod_{mn \atop d_{mn} = 0} \left[ \frac{\mu^4}{M^2 M_p^2} \left( \frac{H_s}{M} \right)^{- 2}  \right]^{V_{mn}} \times \prod_{r \atop d_r = 0}  \left[ \frac{\mu^4}{M^2 M_p^2} \left( \frac{H_s}{M} \right)^{- 2}  \right]^{V_r}
    = \prod_{d_i = 0} \left( \frac{\mu^4}{M_p^2 H_s^2} \right)^{V_i} \,,
\ee
where the index $i$ runs over both $(mn)$ and $r$ types of vertices. For inflationary applications where the scalar potential dominates the gravitational response we know $\mu^2 \sim H M_p$ and so this factor becomes
\be
     \prod_{d_i = 0} \left( \frac{H^2}{H_s^2} \right)^{V_i} = \prod_{d_i = 0} c_s^{2V_i} \,,
\ee
as claimed.

\subsection{Limiting cases}

Eq.~\pref{MasterBresult} has a number of limits that can be checked against other calculations. 

\subsubsection*{Case 1: $c_s \to 1$}

First, if $c_s \to 1$ and if no $A$ fields are in a graph\footnote{An absence of $A$ fields in the graph is required because the small-$c_s$ limit makes the expansion in powers of $A$ be dominated by a quadratic action coming from the square-root term of the DBI action, whereas in the relativistic problem it comes from the Einstein-Hilbert action (as does the quadratic action of $B$ above).} then the result should go over to the power-counting expression for relativistic scalars coupled to gravity, as computed in \cite{InfPowerCount, ABSHclassic}.   In the $c_s \to 1$ and $\cE^A = N^A_{mn} = N^A_r = 0$ limits and dropping the square-root interactions parameterized by $\mathfrak{a}_m$, the above expression becomes
\be
 \cB_{\cE}(H) \to \frac{M_p^2}{H^2} \,  \left( \frac{H}{4\pi M_p} \right)^{2L } \left( \frac{ H^2}{M_p} \right)^{\cE}  \prod_n \left[\mathfrak{c}_n   \left( \frac{H}{M} \right)^{d_{n} - 2}   \right]^{V_n} \,,
\ee
which uses $H_s \to H$ as $c_s \to 1$. Here $\cE$ counts all external lines (scalar and metric) and $n$ denotes the generic sum over all possible effective interactions. This agrees with the results of \cite{ABSHclassic} once it is recognized that the dimensionless effective couplings $\mathfrak{c}_n$ contain a factor of $(M/M_p)^2$ for all interactions with more than 2 derivatives. (This factor is required if the action descends from a derivative expansion of the form \pref{VanillaEFT} with higher derivatives suppressed solely by powers of $M$.)

\subsubsection*{Leading contributions to 2-point functions}

Another useful limit is the lowest-order contributions to the two-point functions for scalar fluctuations,\footnote{At lowest order the tensor-tensor correlations do not `see' the DBI action at all and so do not depend on $c_s$. For example the leading piece predicted for $\langle E \, E \,\rangle$ only come from the Einstein-Hilbert term and so agree with the standard expressions.} since these are explicitly known \cite{BFM, GM, Bean:2008ga}. 

The quantity of practical interest in this case is the curvature perturbation on super-Hubble scales, $\zeta$, evaluated for physical momenta equal to the sound horizon, $q/a \sim H_s \sim H/c_s$. In our gauge $\zeta$ arises as a linear combination of the fluctuations $\phi$ and $B$, 
\be
  \zeta \sim \frac{B}{M_p} + \frac{H  \phi}{\dot \varphi}
\ee
up to order-unity factors \cite{BFM, BaumannTASI}. Consequently $\langle \zeta \, \zeta \rangle$ arises as a linear combination of $\langle \phi \,\phi \rangle$ and $\langle B \, B  \rangle$ and so on. Since, during inflation, $H \sim \mu^2/M_p$ and $\dot \varphi \sim \Lambda^2 \sqrt{1-c_s^2} \sim \Lambda^2$ it follows that 
\be
  \frac{H \phi}{\dot \varphi} \sim  \frac{\mu^2}{\Lambda^2} \left( \frac{\phi}{M_p} \right) 
\ee
where $\mu^2/\Lambda^2 \ll 1$.  

Specializing \pref{MasterBresult} to $\cE^\phi = 2$ and $\cE^A=\cE^B = 0$ or to $\cE^B=2$ and $\cE^\phi = \cE^A = 0$, and so on, one finds the leading contribution corresponds to $L = 0$ ({\it i.e.} tree level or classical contribution) with no insertions of interactions ({\it i.e.} $V_{mn} = V_r = 0$ for all $(mn)$ and $r$), giving
\be
 \langle \phi \, \phi \rangle \sim c_s^2 \langle \psi \, \psi \rangle \sim H^2 \quad \hbox{and} \quad
 \langle B \, B \rangle \sim c_s \langle \hat B \, \hat B \rangle \sim \frac{H^2}{c_s} \,,
\ee
corresponding to the free correlation functions $\langle \psi \,\psi \rangle \sim \langle \hat B\, \hat B \rangle \sim H_s^2 \sim H^2/c_s^2$. These show that it is the $\langle B \, B \rangle$ correlation that dominates in the curvature fluctuation, leading to the estimate\footnote{In addition to the $c_s$-dependence found by the power-counting arguments given here, eq.~\pref{zetazetasize} also quotes the leading slow-roll dependence, as obtained from power-counting in \cite{ABSHclassic} (and in agreement with explicit calculations).}
\be \label{zetazetasize}
   \langle \zeta \, \zeta \rangle \sim \frac{H^2}{ \epsilon \, M_p^2 c_s^2} 
\ee
in agreement with the $c_s$-dependence of more detailed calculations \cite{BFM, GM, Bean:2008ga, Kinney:2007ag}. 

\subsection{A dangerous loop}
\label{dangerousloop}

We close by circling back to an issue discussed earlier: graphs with graviton loops can give less suppression by powers of $c_s$ than is predicted by \pref{MasterBresult}, due to the inability to scale $c_s$ out of both the scalar and metric propagators simultaneously.  

\begin{figure}[!ht]
\begin{centering}
\includegraphics[scale=0.5]{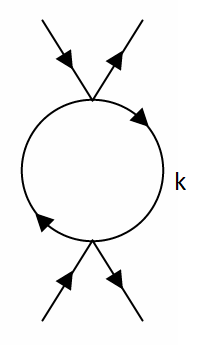}
\par\end{centering}
\caption{\label{fig:temp}Loop diagram generated by the $\frac{c_{s}}{\Lambda^4} \, E^2 (\psi')^2$ vertex}
\end{figure}

The issue here is most easily expressed using an explicit example. To this end consider one of the tensor-scalar interactions that lies buried in the square-root part of the DBI action, 
\be
   \mathfrak{L}_\eff \supset \frac{ 1}{M_p^2} \, E_{ij} E^{ij} ({\psi')^{2}} \simeq \frac{c_{s}}{M_p^2}\, \hat E_{ij} \hat E^{ij} ({\psi')^{2}} \,,
\ee
coming from the DBI square-root term. When used in a loop this interaction generates a correction to the effective $(\psi')^4$ interaction, through the Feynman diagram of figure \ref{fig:temp}, where it is the tensor mode $E_{ij}$ that circulates within the loop.   

At face value the dimensional power-counting arguments given above would predict that the correction to the coefficient of the $(\psi')^4$ term obtained from this graph is suppressed by two powers of $c_s$, since
\be
  \delta \mL_\eff \sim \frac{c_s^2}{(4\pi)^2 M_p^4} \, (\psi')^4 \,,
\ee
with the loop integral contributing a dimensionless logarithmic coefficient. This naive counting over-estimates the suppression by $c_s$, however, as is now shown in more detail. 

In the momentum space the loop integral arising within this type of graph is 
\be 
    \int \frac{\exd^{3}K \exd k^{0}}{(2\pi)^{4}} \frac{c_{s}^{2}}{(c_{s}^{2}(k^{0})^{2}-K^{2})^{2}} ,
\ee
where $K = |\Vec{k}|$. The $c_{s}$-dependence of this integral can be made more explicit by performing a change of coordinates, $k^0 \to u$ with $u := c_{s} k^{0}$. This substitution leads to 
\be \label{eq: zadnja}
    \int \frac{d^{3}K du}{(2\pi)^{4}} \frac{c_{s}}{(u^{2}-K^{2})^{2}} \,.
\ee
Equation \eqref{eq: zadnja} shows that the loop contribution is suppressed by one less power of $c_s$ than would have been obtained from our power-counting estimate. As mentioned earlier, the answer differs from the estimate because the integral is dominated by configurations for which $k^0 \sim K/c_s$, in which case it is wrong to regard  a term like $c_s^2 (E')^2$ to be a small perturbation relative to $(\nabla E)^2$. 

Although the suppression by powers of $c_s$ are not as big as expected, the graph is still suppressed by positive powers of $c_s$. Is this always the case, or can this kind of loop effect undermine the validity of a systematic small-$c_s$ expansion? We now argue that the small-$c_s$ expansion nonetheless survives.

To see why consider a graviton loop that appears somewhere within a more complicated graph. Suppose also that this loop has $N_\ssV \ge 1$ vertices on it, to each of which at least two graviton legs must attach. The argument above shows that for each loop of this type there is an enhancement of the graph by a factor of $1/c_s$ relative to the power-counting arguments given above. But these same power-counting arguments assign a factor of $\sqrt{c_s}$ each time that a field $E_{ij}$ appears\footnote{The power-counting of the $E_{ij}$ couplings proceeds along much the same lines as was found for the field $B$ above.} in a vertex, and so all of the vertices along the graviton loop must contribute a factor of $c_s^{N_\ssV}$, since there are at least two graviton legs for each vertex. Because $N_\ssV \geq 1$ these vertex factors of $c_s$ are always able to overwhelm the $1/c_s$ coming from the loop integration, always leaving nonegative powers of $c_s$. 

\section{Summary and conclusions}
\label{sec:Conclusions}

In summary, we have shown how the square-root structure of the action for DBI models can allow a controlled semiclassical EFT expansion to exist in the regime where $\dot\varphi^2$ is not small and $c_s \ll 1$. This is at first sight surprising, since most of our understanding of quantum effects in gravity rely on a low-energy derivaive expansion, while a large deviation in $1-c_s$ normally requires large derivatives.  

The controlled expansion is nevertheless possible in this regime because of the new small parameter $c_s \ll 1$. It is this small quantity that makes the calculation usefully constrained, since all non-DBI higher-derivative corrections come suppressed by positive powers of $c_s$. The issues that arise strongly resemble those that come up in NRQED, due to the nonrelativistic dispersion relation satisfied by the DBI scalar. The power-counting becomes disrupted by the widely different propagation speeds experienced by the graviton and the DBI scalar, though not to the point where the expansion in $c_s$ and powers of $H/M_p$ breaks down.   

\section*{Acknowledgements}
We thank Ana Ach\'ucarro, Subodh Patil, Eva Silverstein, Andrew Tolley, and David Tong for helpful discussions about power-counting and small-$c_s$ models, and in particular Peter Adshead, Rich Holman and Sarah Shandera for early discussions of this topic at the Banff International Research Station. We also thank Supranta S.~Boruah and Luk\'a\v{s} Gr\'af, who participated in early stages of this project. This work was partially supported by funds from the Natural Sciences and Engineering Research Council (NSERC) of Canada. Research at the Perimeter Institute is supported in part by the Government of Canada through NSERC and by the Province of Ontario through MRI.

\end{document}